\documentclass[final,3p,11pt]{elsarticle}

\usepackage{hyperref}
\usepackage{graphicx,wrapfig}
\usepackage{subfigure}
\usepackage{caption}
\usepackage{amssymb}
\usepackage{amsmath,mathrsfs}
\usepackage{bm}
\usepackage{url}
\usepackage[dvipsnames]{xcolor}
\usepackage{lipsum}
\usepackage{csquotes}
\usepackage{verbatim}
\usepackage{mathtools}
\usepackage{listings}
\usepackage[export]{adjustbox}
\usepackage{tikz}

\numberwithin{equation}{section}

\newcommand{\change}[1]{{\color{black}#1}}

\newcounter{numfunc}
\newenvironment{nfunc}[1]{%
  \stepcounter{numfunc}
  \expandafter\xdef\csname#1\endcsname{\thenumfunc}%
  \quote \thenumfunc: \ignorespaces}{\unskip\endquote}
\newcommand\funcref[1]{\csname#1\endcsname}

\usepackage{float}

\makeatletter
\def\VerbLB{\FV@Command{}{VerbLB}}
\begingroup
\catcode`\^^M=\active%
\gdef\FVC@VerbLB#1{%
  \begingroup%
    \FV@UseKeyValues%
    \FV@FormattingPrep%
    \FV@CatCodes%
    \def^^M{ }%
    \catcode`#1=12%
    \def\@tempa{\def\FancyVerbGetVerb####1####2}%
    \expandafter\@tempa\string#1{\mbox{##2}\endgroup}%
    \FancyVerbGetVerb\FV@EOL}%
\endgroup
\makeatother

\allowdisplaybreaks

\begin{document}

\begin{frontmatter}

\title{SFQEDtoolkit: a high-performance library for the accurate modeling of strong-field QED processes in PIC and Monte Carlo codes}

\author[MPIK]{Samuele Montefiori}
\author[MPIK]{Matteo Tamburini\corref{author}}

\cortext[author] {Corresponding author.\\\textit{E-mail address:} matteo.tamburini@mpi-hd.mpg.de}

\address[MPIK]{Max-Planck-Institut f{\"u}r Kernphysik, Saupfercheckweg 1, D-69117 Heidelberg, Germany}

\begin{abstract}
Strong-field quantum electrodynamics (SFQED) processes are central in determining the dynamics of particles and plasmas in extreme electromagnetic fields such as those present in the vicinity of compact astrophysical objects or generated with ultraintense lasers. SFQEDtoolkit is an open source library designed to allow users for a straightforward implementation of SFQED processes in existing particle-in-cell (PIC) and Monte Carlo codes. Through advanced function approximation techniques, high-energy photon emission and electron-positron pair creation probability rates and energy distributions are calculated within the locally-constant-field approximation (LCFA) as well as with more advanced models [Phys. Rev. A \textbf{99}, 022125 (2019)]. SFQEDtoolkit is designed to provide users with high-performance and high-accuracy, and neat examples showing its usage are provided. In the near future, SFQEDtoolkit will be enriched to model the angular distribution of the generated particles, i.e., beyond the commonly employed collinear emission approximation, as well as to model spin and polarization dependent SFQED processes. Notably, the generality and flexibility of the presented function approximation approach makes it suitable to be employed in other areas of physics, chemistry and computer science.
\end{abstract}


\end{frontmatter}

\section{Introduction}

Since the invention of chirped pulse amplification, the continuous technological progress in high-power laser systems has opened up the possibility of accessing ultrastrong electromagnetic fields in the laboratory. Peak laser intensities of the order of $10^{23}$~W/cm$^2$ have been recently achieved with multi-petwatt laser systems~\cite{yoonO21}, while 10~PW lasers are expected to become soon available at the Extreme Light Infrastructure (ELI)~\cite{eliURL} and in several other facilities worldwide~\cite{dansonHPLSE19}. This new experimental capability has driven substantial theoretical and experimental effort in exploring the strong-field frontier of quantum electrodynamics (QED) from atoms~\cite{mockenJPC04, mockenCPC05} to plasmas~\cite{marklundRMP06, dipiazzaRMP12, zhangPoP20, fedotovXXX22}. Indeed, first experiments coupling ultrarelativistic electron beams with intense lasers have recently provided the first evidence of radiation reaction effects and of their quantum nature~\cite{poderPRX18, colePRX18}. In parallel, progress in accelerator technology with the possibility of generating high-current and strongly focused ultrarelativistic lepton beams has provided a novel route to investigate SFQED in beam-beam and in beam-plasma interaction~\cite{yakimenkoPRL19, delgaudioPRAB19, tamburiniPRD21, sampathPRL21, matheronXXX22}. Further technological development could even allow to probe the supercritical regime of SFQED, where particles experience fields largely exceeding the QED critical one $F_{cr}=m_e^2 c^3/|e| \hbar \approx 1.3 \times 10^{18}\;\text{V/m}$ ($\approx 4.4\times 10^{9}\;\text{T}$) in their rest frame. Here $c$ is the speed of light in vacuum, $\hbar$ the reduced Planck constant, while $m_e$ and $e$ are the electron mass and charge, respectively. 

In relativistic laboratory astrophysics, the study of SFQED processes in plasmas with even the possibility to generate dense electron-positron jets extending over several plasma skin depths~\cite{sarriPRL13, sarriNC15} could enable access to the microphysics that determines both the plasma dynamics and the powerful emission processes thought to occur around compact astrophysical objects such as pulsars and magnetars~\cite{ruffiniPR10, uzdenskyRPP14, ceruttiMNRAS16, philippovPRL20}. This unique opportunities justify the growing interest and effort in studying SFQED effects in extreme-field plasma physics. A prominent example are QED cascades, whose dynamics is essentially determined by the interplay between SFQED and the collective plasma response~\cite{bellPRL08, fedotovPRL10, elkinaPRSTAB11, nerushPoP11, nerushPRL11, ridgersPRL12, bashmakovPoP14, gelferPRA15, grismayerPoP16, grismayerPRE17, tamburiniSR17, jirkaSR2017, sampathPOP18}.

SFQED-dominated plasmas are complex and strongly nonlinear systems. Modeling and gaining insights into the intrinsically multiscale dynamics that governs these complex systems in realistic conditions critically depends on the development of advanced numerical tools and codes that can efficiently run on high-performance computing (HPC) machines. In the last few decades, several Monte Carlo and PIC codes have been developed to study beam physics and classical plasma dynamics~\cite{birdsall-langdon}. Some of these codes have been enriched with SFQED routines that model high-energy photon emission and electron-positron pair creation in strong background electromagnetic fields~\cite{elkinaPRSTAB11, ridgersJCP14, arberPPCF15, gonoskovPRE15, lobetJPCS16, tamburiniSR17, nielPPCF18, fedeliNJP22, guoPRE22}, and merging algorithms~\cite{vranicCPC15, muravievCPC21} have been developed to tackle the excessive or even unfeasible computational demands caused by the rapid and potentially exponential growth of the number of computational particles in SFQED simulations. However, to our knowledge, the standard approach to model probability rates and distributions of SFQED processes in state-of-the-art codes resorts to binary search and interpolation applied to pre-computed lookup tables where the required functions are sampled at a limited number of points logarithmically distributed between a minimum and a maximum that can span several orders of magnitude~\cite{ridgersJCP14, lobetJPCS16, fedeliNJP22}. This approach is motivated by the fact that SFQED rates and distribution functions involve numerically expensive special functions and integrals whose evaluation at runtime would largely exceed the computational cost of PIC routines, therefore rendering SFQED PIC simulations unviable, practically. However, the major shortcomings of this approach are that: (i) it struggles to cover the whole range of parameters. In fact, e.g., a cut-off in the photon spectrum is commonly introduced. (ii) to achieve high accuracy a very high number of sampling points is needed, which noticeably increases the computational cost. It is worth noting that the insufficient resolution of lookup tables may introduce both substantial systematic errors in probability rates (see section~\ref{ph_em}) and artifacts such as stairlike structures in the high-energy region of the photon spectrum~(see Ref.~\cite{guoPRE22} and section~\ref{ph_em}). The above shortcomings can unacceptably affect the reliability of simulation results.

Here we introduce SFQEDtoolkit, a suite of modules designed to implement SFQED processes in existing PIC and Monte-Carlo codes while overcoming the above-mentioned issues. We follow the standard methodology for implementing SFQED processes in a code, which consists of two main steps. First, determine whether the SFQED event is deemed to occur according to its rate, i.e., its probability per unit time. This is a critical step and needs to be highly optimized because SFQED rates are required for each particle and at each timestep, thereby critically affecting both the performance and the accuracy of the simulation. Second, if the event occurs, sample the state after the event from its probability distribution. In the collinear emission approximation, this reduces to calculate the particles' energy after the event.

Note that in the $S$-matrix formalism of SFQED the calculation of the probability involves integrals in time ranging from $-\infty$ to $+\infty$, i.e., one needs to consider the asymptotic states in the past and in the future. The use of rates depending only on the instantaneous value of particles' parameters is possible when SFQED probabilities are formed over a temporal scale much smaller than the scale of variation of electromagnetic fields~\change{(see Refs.~\cite{ritusJSLR85, baierNPB89} for further details)}. This assumption constitutes the core of the widely employed locally-constant-field approximation (LCFA). In the paradigmatic case of photon emission by an electron colliding with a laser pulse, LCFA requires \change{$\xi \equiv a_0 = |e| E/ m_e \omega c \gg 1$} and $\xi^3 \gg \chi_e$, where $\chi_e = F^*/F_{cr}$ with $F^*$ being the field amplitude in the instantaneous rest frame of the electron, while $E$ and $\omega$ are the laser field amplitude and frequency, respectively~\cite{ritusJSLR85, baierNPB89, dinuPRL16}. \change{Physically, the above conditions imply that SFQED probabilities receive most of their contribution from a region of a particle's trajectory that is small compared to the scale of variation of the background electromagnetic fields, and that most of the radiated photon spectrum is well approximated by the LCFA photon spectrum~\cite{dipiazzaPRA18, dipiazzaPRA19}.} \change{Interestingly, however, even when the above conditions are fulfilled the LCFA photon spectrum differs quantitatively and qualitatively from the exact one for relatively low emitted photon energies  $\varepsilon_\gamma \lesssim (\chi_e/\xi^3) \varepsilon_e$}, where $\varepsilon_\gamma$ and $\varepsilon_e$ are the photon and the electron energy, respectively~\cite{dipiazzaPRA18}. To properly model this region of the photon spectrum one has to resort to more advanced techniques~\cite{dipiazzaPRA19, wistisenPRR19, wistisenPRD19b, heinzlPRA20, nielsenPRD22}. On the one hand, some of these techniques accurately reproduce the interference features of the emitted photon spectrum predicted by SFQED. On the other hand, most of them are suitable only for specific cases such as the interaction of particles with a plane-wave pulse or with a crystal. In SFQEDtoolkit we implement the technique described in Ref.~\cite{dipiazzaPRA19}. \change{Albeit this technique cannot exactly reproduce the SFQED spectrum for all photon energies, it provides a several orders of magnitude better approximation than the LCFA in the low energy part of the photon spectrum while retaining the advantages of the LCFA. Namely, the high energy part of the photon spectrum corresponds to the SFQED spectrum, it depends only on the local instantaneous value of a particle's parameters, and is suitable for background electromagnetic fields with arbitrary spacetime structure such as those occurring in PIC simulations.}

SFQEDtoolkit can be used as a black box, \change{in which case the user can directly refer to \ref{user}, which summarizes the currently available routines and the essential steps for a straightforward implementation of SFQEDtoolkit in a code}. More detailed instructions, updates and neat examples of \change{the usage of SFQEDtoolkit} are provided in \texttt{C++} and \texttt{Fortran} \change{in the ``example\_cpp'' and ``example\_fortran'' folders in the GitHub repository} at \url{https://github.com/QuantumPlasma/SFQEDtoolkit}.

\section{Rates and cumulative probabilities} \label{probabs}

The emission of a photon by an electron and the conversion of a photon into an electron-positron pair in a background electromagnetic field are stochastic events. \change{Under suitable conditions such as those of the LCFA, the first order $S$-matrix contribution to the probability of an event in a short temporal interval $dt$ is much smaller than unity and the probability of the process is linearly proportional to $dt$~\cite{ritusJSLR85, baierNPB89}.} This is at the heart of the local probability method, where a stochastic event is deemed to occur if $R(t) dt > r$. Here $R(t)$ is the rate of the process at time $t$, and $0<r<1$ is a uniformly distributed random number. However, in general, the emission probability of several photons over an arbitrary interval, is not known. 

In this section we present the theory that justifies one of the broadly used stochastic techniques, namely, the optical depth method (see also the statistical treatment in Ref.~\cite{tamburiniPRD21}). We then compare its advantages and disadvantages with respect to the local probability method. We start by introducing the probability $P_n(0, t)$ that an electron emits $n$ photons in an arbitrary temporal interval $(0, t)$. In the following, we assume that:
\begin{enumerate}
\item the probability of emission in a small temporal interval $dt$ depends only on the instantaneous value of physical parameters at $t$;
\item for small $dt$ the probability of multiple $n-$photon emissions is negligibly small. Hence, the only possible event in $dt$ is a single photon emission;
\item the electron motion is quasiclassical so, between two emission events, its trajectory is well defined and obtained by solving the Lorentz equation of motion.
\end{enumerate}
The probability of no emission $P_0(0, t+dt)$ in an interval $(0,t+dt)$ can be written in terms of the probability of no emission $P_0(0,t)$ in $(0,t)$ and the probability of no emission $P_0(t,t+dt)$ in $(t,t+dt)$. In fact, from the first assumption $P_0(0,t)$ and $P_0(t, t+dt)$ are independent. Thus, $P_0(0, t+dt) = P_0(0, t) P_0(t, t+dt)$. In addition, from the second assumption $P_n(t, t + dt) = 0$ for $n>1$, which in turn implies that $P_0(t, t+dt) = 1 - P_1(t, t +dt)$. Since the probability of emission in a sufficiently small time interval $dt$ is $P_1(t, t + dt) = R_{pe}(t) dt$, where $R_{pe}(t)$ is the rate of photon emission at time $t$, one gets the differential equation
\begin{equation} \label{diff_eq_1}
\frac{dP_0 (0, t)}{dt} = - R_{pe}(t)P_0 (0, t),
\end{equation}
whose solution after setting \change{$P_0 (0, 0) = 1$} as initial condition is
\begin{equation} \label{diff_eq_solution_1}
P_0 (0, t) = \exp{\Bigl(- \int_{0(\Gamma_0)}^{t} R_{pe}(\tau)d\tau\Bigr)}.
\end{equation}
In Eq.~\eqref{diff_eq_solution_1}, $\Gamma_0: t \to [\mathbf{x}(t),\mathbf{p}(t)]$ denotes the curve representing the electron trajectory \change{in phase space} which, according to the third of the above hypothesis, is well defined and determined by the Lorentz equation of motion. We stress that, in general, if $\Gamma_a$ and $\Gamma_b$ are two different trajectories then
\begin{equation}
\int_{0(\Gamma_a)}^{t} R_{pe}(\tau)d\tau \neq \int_{0(\Gamma_b)}^{t} R_{pe}(\tau)d\tau.
\end{equation}
From \eqref{diff_eq_solution_1}, the cumulative probability of any event $P_{ae}(0, t)$ in $(0,t)$ is 
\begin{equation} \label{any_event}
P_{ae}(0, t) = \sum_{i = 1}^\infty P_i(0, t) = 1 - \exp{\Bigl(- \int_{0(\Gamma_0)}^{t} R_{pe}(\tau)d\tau\Bigr)},
\end{equation}
Equation~\eqref{any_event} is at the heart of the optical depth method, where a stochastic event is deemed to occur if $P_{ae}(0, t) > r$. In practice, every particle is initialized with an optical depth $\tau_0 = - \ln(1 - r)$ obtained by inverting $P_{ae}(0, t) = r$. The event occurs when $\tau_e(t) > \tau_0$, where $\tau_e(t) = \int_{0(\Gamma_0)}^{t} R_{pe}(\tau)d\tau$ is calculated along the trajectory $\Gamma_0$ obtained from the Lorentz equation. Typically, first order Eulerian integration $\tau_e(t+\Delta t) = \tau_e(t) + R_{pe}(t) \Delta t$ is used in PIC codes, where $\Delta t$ is the timestep (see, e.g., Refs.~\cite{ridgersJCP14, fedeliNJP22}). Notice that, instead of storing both $\tau_0$ and  $\tau_e(t)$, to reduce memory requirements one can equivalently store only $\tau_d(t) = \tau_0 - \int_{0(\Gamma_0)}^{t} R_{pe}(\tau)d\tau$ and check when $\tau_d(t) < 0$. When an event occurs, $\tau_0$ is then re-initialized by using a new random number. 

Although the local and the optical depth method are equivalent, each presents advantages and disadvantages. The main advantage of using an optical depth method is that random numbers are needed only when particles are initialized or after an event occurs. The need of storing the $\tau_d(t)$ of each particle is a relatively minor drawback, and this method may provide better performance when $R_{pe}(\tau)$ can be efficiently calculated. By contrast, the local method can be more efficient when $R_{pe}(\tau)$ is computationally expensive. In this case the local method allows one to use an acceptance-rejection technique. \change{In fact, one can drastically reduce the number of required evaluations of $R_{pe}(\tau)$ by introducing a computationally cheaper $\tilde{R}_{pe}(\tau)$ such that $\tilde{R}_{pe}(\tau) > R_{pe}(\tau)$. Thus, $\tilde{R}_{pe}(\tau) \Delta t > r $ is a necessary but not sufficient condition for $R_{pe}(\tau) \Delta t > r$, $r$ being the same random number. If in the majority of cases $\tilde{R}_{pe}(\tau)$ closely follows $R_{pe}(\tau)$, one can noticeably reduce the number of evaluations of $R_{pe}(\tau)$ by first checking the necessary condition $\tilde{R}_{pe}(\tau) \Delta t > r$, and only in the rare cases where $\tilde{R}_{pe}(\tau) \Delta t > r$, then also $R_{pe}(\tau)$ is evaluated and the condition $R_{pe}(\tau) \Delta t > r$ is checked to decide whether an event is deemed to occur at the considered timestep. Since the rate must be calculated for each particle at each timestep, and since irrespective of the adopted local or optical depth method for good numerical accuracy the timestep must be chosen such that $R_{pe}(\tau) \Delta t \ll 1$, the computational cost of evaluating $R_{pe}(\tau)$ may significantly affect the simulation's performance. A prominent example where the above-mentioned technique is implemented and the local method provides a simpler and more efficient strategy than the optical depth method is discussed in section \ref{blcfa}, where we detail the implementation of photon emission beyond the LCFA model. In fact, in the beyond the LCFA model $\tilde{R}_{pe}(\tau)$ reduces to a simple one parameter function, while $R_{pe}(\tau)$ is a relatively complicated multivariate function.}

It is worth determining also $P_1 (0, t), \dots, P_n (0, t)$ when possible, \change{i.e., when there is essentially a single possible trajectory in phase space (see below).} We begin by considering the probability of a single photon emission $P_1 (0, t + dt)$ in the interval $(0, t + dt)$. In this case only two mutually exclusive channels are possible: either no emission in $(0, t)$ followed by one photon emission in $(t, t + dt)$, or one emission in $(0, t)$ followed by no emission in $(t + dt)$. The corresponding equation is
\begin{equation} \label{prob_single_em}
P_1 (0, t + dt) = P_0 (0, t)P_1 (t, t + dt) + P_1 (0, t)P_0 (t, t + dt).
\end{equation}
Following the same reasoning as above, we have
\begin{align}
P_1(t,t+dt)& = R_{pe}[\varepsilon_{e,0}(t),\chi_{e,0}(t)]dt \equiv R_{pe,0}(t)dt 
\label{prob_s_em}\\
P_0(t,t+dt)& = 1 - R_{pe}[\varepsilon_{e,1}(t),\chi_{e,1}(t)]dt \change{\equiv 1 - R_{pe,1}(t)dt}
\label{prob_z_em}
\end{align}
where for clarity the dependence of $R_{pe}$ on the electron energy and quantum parameter have been explicitly reported, and the subscript 0 (1) denotes quantities from the trajectory \change{in phase space} without (with) photon emission. By replacing Eq.~\eqref{diff_eq_solution_1}, Eq.~\eqref{prob_s_em} and Eq.~\eqref{prob_z_em} into Eq.~\eqref{prob_single_em} we obtain the differential equation
\begin{align} \label{replace}
\frac{dP_1(0, t)}{dt} + R_{pe,1}(t) P_1(0, t) = R_{pe,0}\exp{\Bigl(- \int_{0(\Gamma_0)}^{t} R_{pe,0}(\tau)d\tau\Bigr)},
\end{align}
with the initial condition $ P_1(0, 0) = 0$. \change{Along a specific trajectory in phase space,} Eq.~\eqref{replace} can be solved with the integrating factor method. However, in practice, neither $\varepsilon_{e,1}(t)$, $\chi_{e,1}(t)$ nor \change{the time of emission are uniquely defined. In fact, an infinite number of $\varepsilon_{e,1}$, $\chi_{e,1}$ and emission instants are possible, such that an infinite number of $\Gamma_1$ need to be considered, and the integral along a particular $\Gamma_1$ does not give the probability of one emission in $(0, t)$. In the limiting case either of $R_{pe,1}(t) = 0$ or of soft photon emissions, where the electron energy and its dynamics are basically unaffected by the emission events such that $R_{pe,1}(t) \approx R_{pe,0}(t)$, there is a single phase space trajectory, and one can determine the corresponding total probabilities.} In the \change{$R_{pe,1}(t) = 0$} case one immediately gets
\begin{align} \label{solution_simp_pre}
P_1 (0, t) = 1 - \exp{\Bigl(- \int_{0(\Gamma_0)}^{t} R_{pe}(\tau)d\tau\Bigr)},
\end{align}
which consistently coincides with the total probability of any event in Eq.~\eqref{any_event}. This reasoning applies to photon conversion into an electron-positron pair where, obviously, the assumption that no event is possible after the first one holds rigorously. In the soft-photon emission case one gets
\begin{align} \label{solution}
P_1 (0, t) = \Bigl(\int_{0(\Gamma_0)}^{t} R_{pe}(\tau)d\tau\Bigr)\exp{\Bigl(- \int_{0(\Gamma_0)}^{t} R_{pe}(\tau)d\tau\Bigr)}.
\end{align}
Following the same reasoning that has led to Eq.~\eqref{prob_single_em}, the probability of $n-$photon emissions is
\begin{equation} \label{prob_multi_em}
P_n (0, t + dt) = P_{n-1} (0, t)P_1 (t, t + dt) + P_n (0, t)P_0 (t, t + dt).
\end{equation}
For soft photon emission this leads to the equation
\begin{equation} \label{replace_n}
\frac{dP_n(0, t)}{dt} + R_{pe,0}(t) P_n(0, t) = R_{pe,0}(t) P_{n-1}(0, t),
\end{equation}
whose solution is
\begin{equation} \label{final_sol}
P_n(0, t) =\frac{\Bigl(\int_{0(\Gamma_0)}^{t} R_{pe}(\tau)d\tau\Bigr)^{n}}{n!}\exp{\Bigl(- \int_{0(\Gamma_0)}^{t} R_{pe}(\tau)d\tau\Bigr)}.
\end{equation}
Thus, the probability of emission of $n$ soft photons follows a Poisson distribution, and $\int_{0(\Gamma_0)}^{t} R_{pe}(\tau)d\tau$ is the mean number of emitted photons in the interval $(0, t)$. This result corresponds to the classical limit, where multiple soft photon emissions occur, and each photon carries away a negligible fraction of the electron energy~\cite{glauberPR51, dipiazzaPRL10}.

\section{Methodology of implementation } \label{impl}

SFQEDtoolkit aims at computing rates and probability distributions of SFQED processes with better than 0.1\% accuracy throughout the domain of interest with performance better than or similar to coarse lookup tables. The open-source PIC code Smilei with its default 256 points\footnote{256 points in 1D and $256 \times 256$ for 2D tables.} and 1024 points\footnote{1024 points in 1D and $1024 \times 1024$ for 2D tables.} precomputed tables is used as a benchmark~\cite{derouillatCPC18}. The methodology employed to implement SFQED processes in SFQEDtoolkit chiefly resorts to a combination of: (i) function approximation with Chebyshev polynomials; (ii) asymptotic expansions; (iii) variable and function transformation. 

We start by briefly describing Chebyshev polynomials and discussing a few properties that render them an ideal basis for approximating smooth functions. Chebyshev polynomials are defined via the recurrence relation
\begin{equation} \label{cheb_rec_rel}
T_{n+1}(x) = 2xT_n(x) - T_{n-1}(x), \quad \text{with} \quad T_0(x)=1, \quad  T_1(x)=x,
\end{equation}
and form an infinite-dimensional basis of functions orthogonal with respect to the weight $1/\sqrt{1-x^2}$ and defined in the interval $-1 \leq x \leq 1$, i.e.,
\begin{align} \label{cont_ortho}
\int^{+1}_{-1}{\frac{T_i(x) T_j(x)}{\sqrt{1-x^2}}} = \left\{
    \begin{aligned}
        0 & \quad & i \neq j & \\
        \pi/2 & \quad & i = j \neq 0 & \\
        \pi & \quad & i=j=0 &
    \end{aligned}
    \right. 
\end{align}
An explicit expression of Chebyshev polynomials is $T_n(x)=\cos(n \arccos x)$, which highlights their relation with the discrete Fourier transform. A continuous function $f(x)$ can be approximated as
\begin{equation} \label{f_approx}
f(x) \approx \sum_{i=0}^{N} c_i T_{i}(x),
\end{equation}
where $c_i$ are the coefficients of the expansion and $N$ is the order of the expansion, i.e., the higher order polynomial in the expansion. The power of function approximation with Chebyshev polynomials resides in: 
\begin{enumerate}
\item The coefficients $c_i$ can be easily calculated by exploiting the orthogonality relations of Chebyshev polynomials (see Eq. \eqref{cont_ortho} or, e.g., Ref.~\cite{pressNR3rd} for the discrete orthogonality relationships). To calculate the coefficients used by SFQEDtoolkit, we have developed a parallel code that efficiently computes Chebyshev coefficients of an arbitrary multivariable 1D, 2D, and 3D function defined in a finite range of its parameters until a predefined accuracy is achieved. Note that a variable $x$ in the interval $a<x<b$ can be mapped on a variable $X$ in the interval $-1<X<1$ via $X = (2 x - a - b)/(b-a)$. The inverse transformation is $x = [(b-a)X + a + b]/2$.
\item A Chebyshev approximation is nearly the same as the minimax polynomial, i.e., the polynomial which has the smallest maximum deviation from the true function $f(x)$ among all polynomials of the same degree. However, unlike the exact minimax polynomial, its coefficients can be readily computed.
\item The error is almost uniformly distributed over the entire approximation range.
\item The magnitude of the first neglected Chebyshev coefficient provides a good estimate of the error of the approximation. This is due to $T_n(x)=\cos(n \arccos x) \leq 1$.
\item For smooth functions, its Chebyshev expansion is rapidly convergent. This allows to condense most of the information on the function in relatively few coefficients that can fit in a modern CPU cache. Indeed, according to Ref.~\cite{pressNR3rd}:
\begin{displayquote}
Any function that is bounded on the interval [of the Chebyshev polynomials] will have a convergent Chebyshev approximation as [the order] $N\to\infty$, even if there are nearby poles in the complex plane. For functions that are not infinitely smooth, the actual rate of convergence depends on the smoothness of the function: the more derivatives that are bounded, the greater the convergence rate. For the special case of a $C^\infty$ function, the convergence is exponential.
\end{displayquote}
\item Due to its recurrence relation in Eq.~\eqref{cheb_rec_rel}, a Chebyshev expansion can be efficiently evaluated with the Clenshaw's recurrence formula~\cite{clenshawMC55}. This enables one to evaluate the expanded function using only the coefficients of the expansion $c_i$, without evaluating the actual polynomials $T_i(x)$. Moreover, Clenshaw's recurrence formula guarantees both efficiency and numerical stability against roundoff errors (see~\ref{clensh});
\item A \change{generic} multivariable function $f(x,y)$ can be expanded as
\begin{equation} \label{doublef_expans_cheb}
f(x,y) \approx \sum_{i=0}^{N}\sum_{j=0}^{M} c_{ij} T_{i}(x)T_{j}(y).
\end{equation}
If $f(x,y)$ needs to be evaluated multiple times, e.g., at several different $x$ but with the same $y$, its computational cost can be greatly reduced. In fact, Eq.~\eqref{doublef_expans_cheb} can be rewritten as
\begin{equation} \label{doublef_expans_cheb_2}
f(x,y) \approx  \sum_{i=0}^{N} \left[\sum_{j=0}^{M} c_{ij} T_{j}(y) \right] T_{i}(x) = \sum_{i=0}^{N} d_i(y) T_{i}(x).
\end{equation}
Clenshaw's recurrence is first used to compute the $N+1$ coefficients $d_i(y)$. Then, it is further used for evaluating the function in $x$ by employing $d_i(y)$ as coefficients. After the first iteration, the computational cost for evaluating the function at any $x$ with same $y$ is therefore reduced to that of a single variable function with $N+1$ coefficients instead of a two variable function with $(N+1)\times(M+1)$ coefficients. This is highly beneficial, for example, when employing root finding routines.
\end{enumerate}
The above properties are systematically used in SFQEDtoolkit. \change{Given a generic function of one $f(x)$ or two $f(x,y)$ arbitrary variables $x$ and $y$ that we want to approximate with a desired accuracy, we divide the interval of definition of $x$ and $y$ into smaller intervals, and in each subinterval of $x$, $y$ the function is approximated with a suitably chosen approximation method. Even when the same approximation method, e.g., a Chebyshev expansion, is used for two distinct subintervals of definition \change{of }the same function, this can result in a substantial reduction of the number of coefficients required to achieve the desired accuracy. At runtime, the microprocessor selects the relevant subinterval among those possible depending on the actual value of the variables, which is called branching. Since modern microprocessors are also pipelined, which means that they can execute multiple code instructions in a single cycle, branching creates a bottleneck. For this reason modern microprocessors have branch predictor capabilities, namely, they select the most likely subinterval of values in a repeated cycle. Thus, a correct prediction of the required branch at runtime may significantly improve the efficient execution of the code. One can take advantage of these capabilities by judiciously choosing the subintervals in which the function is approximated, such that a single subinterval is likely to be most frequently required.}

In the following, we denote the Chebyshev expansion of a function $f(x)$ as $\mathcal{C}[f(x)]$. As we have emphasized above, the rate of convergence of $\mathcal{C}[f(x)]$ strongly depends on the smoothness of $f(x)$. Thus, before calculating a Chebyshev expansion, it is important to analyze $f(x)$ and determine whether its smoothness can be increased, e.g., via a suitable \change{change of variables}. In addition, it can be simpler and computationally cheaper to use asymptotic expansions in the regions of the domain where the function or its derivatives grow rapidly, or when the domain is unbounded. Similarly, asymptotic expansions are preferable when $f(x)$ is exponentially small, or tends to zero much faster than a polynomial. In fact, in this case $f(x)$ and $\mathcal{C}[f(x)]$ tend to zero at different paces, and the accuracy of the approximation decreases. Alternatively, it may be convenient to calculate the Chebyshev expansion of a function derived from $f(x)$. In this case, at runtime one first computes the derived function, and then obtains $f(x)$ from the the evaluation of the derived function. For example, if $f(x)$ or its derivatives are divergent, then $1/f(x)$ or $f(x)/g(x)$ where $g(x)$ is an easily computable asymptotic function that renders $f(x)/g(x)$ smooth, may have a much more rapidly convergent Chebyshev expansion than $f(x)$. From $\mathcal{C}[1/f(x)]$ or $\mathcal{C}[f(x)/g(x)]$, one can easily obtain $f(x)$ as $(\mathcal{C}[1/f(x)])^{-1}$ or $g(x) \mathcal{C}[f(x)/g(x)]$, respectively. In SFQEDtoolkit, the required Chebyshev coefficients are loaded from text files into memory only once when the library is initialized (see \ref{user}).

\section{Photon emission and pair creation in the LCFA} \label{ph_em}

The differential probability of photon emission per unit time $t$ per unit photon energy $\varepsilon_\gamma$
by an electron or positron with energy $\varepsilon_e$ is~\cite{Baier-book}
\begin{equation} \label{pe_dp}
\frac{d^2W_{pe}}{dtd\varepsilon_\gamma}(\varepsilon_\gamma, \varepsilon_e, \chi_e) =  \frac{\alpha m_e^2 c^4}{\sqrt{3}\pi\hbar\varepsilon_e^2}\frac{1}{(1+u)}\Biggl[\bigl[1 + (1 + u)^2\bigr] \mathrm{K}_{\frac{2}{3}}\Bigl(\frac{2u}{3\chi_e}\Bigr) -(1+u) \int_{\frac{2u}{3\chi_e}}^{\infty}\mathrm{K}_{\frac{1}{3}}\bigl(y\bigr)dy\Biggr],
\end{equation}
where $\alpha = e^2/\hbar c \approx 1/137.036$ is the fine structure constant, $u = \varepsilon_\gamma /(\varepsilon_e - \varepsilon_\gamma)$, and $\mathrm{K}_\nu (x)$ are the modified Bessel functions of the second kind. Analogously, the differential probability of electron-positron pair creation per unit time per unit electron energy from a photon with energy $\varepsilon_\gamma$ is~\cite{Baier-book}
\begin{equation} \label{pp_dp}
\frac{d^2W_{pp}}{dtd\varepsilon_e}(\varepsilon_e, \varepsilon_\gamma, \chi_\gamma) =  \frac{\alpha m_e^2 c^4}{\sqrt{3}\pi\hbar\varepsilon_{\gamma}^2}\Biggl[\frac{\varepsilon_e^2 + \varepsilon_p^2}{\varepsilon_e \varepsilon_p}\mathrm{K}_{\frac{2}{3}}\bigl(\eta\bigr) + \int_{\eta}^{\infty}\mathrm{K}_{\frac{1}{3}}\bigl(y\bigr)dy\Biggr],
\end{equation}
where $\eta = 2 \varepsilon_\gamma^2 / (3 \varepsilon_e \varepsilon_p \chi_\gamma)$, and $\varepsilon_p = \varepsilon_\gamma - \varepsilon_e$ is the positron energy. The electron/photon quantum nonlinearity parameter $\chi_{e/\gamma}$ as computed in SFQEDtoolkit by calling either the function \texttt{compute\_chi\_with\_vectors} or the function \texttt{compute\_chi\_with\_components} (see \ref{user} for details) is~\cite{Baier-book}
\begin{equation} \label{chi_exact}
\chi_{e/\gamma} = \frac{\varepsilon_{e/\gamma}}{m_e c^2 F_{cr}} \sqrt{\left(\mathbf{E} + \frac{\mathbf{p}_{e/\gamma} c}{\varepsilon_{e/\gamma}} \times \mathbf{B} \right)^2 - \left(\frac{\mathbf{p}_{e/\gamma} c}{\varepsilon_{e/\gamma}} \cdot \mathbf{E}\right)^2},
\end{equation}
which holds both for electrons and for photons with momentum $\mathbf{p}_{e/\gamma}$ and energy $\varepsilon_{e/\gamma}$ in a background electric $\mathbf{E}$ and magnetic $\mathbf{B}$ field. Eqs.~\eqref{pe_dp}-\eqref{pp_dp} hold if the two field invariants $| \mathbf{E}^2 - \mathbf{B}^2| / F_\text{cr}^2$ and $| \mathbf{E} \cdot \mathbf{B}| / F_\text{cr}^2$ are much smaller than $\text{min}(1, \chi_{e/\gamma}^2)$, and electrons and positrons in the initial and final states are ultrarelativistic $\varepsilon_{e/p}/(m_e c^2) \gg 1$ (see Ref.~\cite{Baier-book}). In fact, in this case the discrete nature of energy levels and the energy associated to the spin degrees of freedom can be neglected, such that a quasiclassical approach is applicable. If background fields are also nearly constant and uniform over the formation length of the process, Eqs.~\eqref{pe_dp}-\eqref{pp_dp} can be used also with time- and space-dependent electromagnetic fields~\cite{ritusJSLR85, Baier-book, dipiazzaPRA18, dipiazzaPRA19}. For completeness, we stress that all currently available SFQED calculations require that perturbation theory in the Furry picture can be used. This assumption was conjectured to break down in the $\alpha \chi^{2/3} \gtrsim 1$ regime, i.e., for $\chi \gtrsim 1600$ (see, e.g., Ref.~\cite{fedotovJPCS17} and references therein). Recent developments have shown that the conjectured breakdown does not occur in general, but its onset depends on the structure of the electromagnetic fields and on the energy of the incoming particles~\cite{podszusPRD19, ildertonPRD19}. Finally, we mention that the assumption of ultrarelativistic particles also allows one to use the collinear approximation, i.e., immediately after the event the momentum of the initial and of the generated particles are aligned. Moreover, the initial (final) spin and polarization degrees of freedom of particles are averaged (summed) in Eqs.~\eqref{pe_dp}-\eqref{pp_dp}. The implementation of the angular distribution of the generated particles beyond the collinear approximation as well as the inclusion of spin-dependent SFQED effects will be presented in a separate publication.

\subsection{Photon emission rate}

\begin{figure}[!tb]
\centering
\includegraphics[width=1.\linewidth]{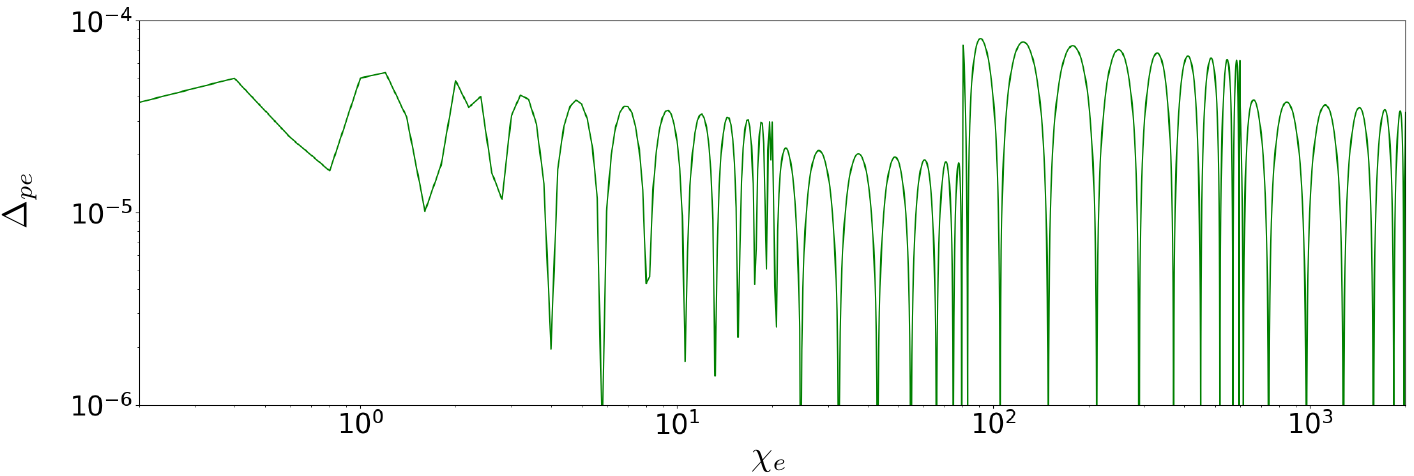}
\caption{Relative accuracy $\Delta_{pe}(\chi_e) = \left| \left\{\tilde{W}_{\text{rad}}(\chi_e) - \mathcal{C}[\tilde{W}_{\text{rad}}(\chi_e) \right\} / \tilde{W}_{\text{rad}}(\chi_e) \right|$ between the analytical and the numerical SFQEDtoolkit photon emission rate (see function \texttt{SFQED\_INV\_COMPTON\_rate} in \ref{user}). Here $\Delta_{pe}(\chi_e)$ is evaluated at $10^4$ evenly spaced points in the interval $0 \leq \chi_e \leq 2000$.} 
\label{phtn_rate_acc}
\end{figure}

\begin{table}
\resizebox{\textwidth}{!}{
\begin{tabular}{ |p{2.8cm}||p{1.5cm}|  }
 \hline
 \multicolumn{2}{|c|}{\texttt{SFQED\_INV\_COMPTON\_rate}} \\
 \hline
 \hspace{1.3cm}$\chi_e$ & \hspace{0.1cm}Method\\
 \hline
 $0 \leq \chi_e < 2$   & C (12) \\
 $2 \leq \chi_e < 20$  & C (11) \\
 $20 \leq \chi_e < 80$ & C (8) \\
 $80 \leq \chi_e < 600$   & C (10) \\
 $600 \leq \chi_e \leq 2000$  & C (7) \\
 \hline
\end{tabular}
\begin{tabular}{ |p{1.5cm}|p{2.6cm}|p{2.7cm}|p{1.5cm}|  }
 \hline
 \multicolumn{4}{|c|}{\texttt{SFQED\_LCFA\_INV\_COMPTON\_PHOTON\_energy}} \\
 \hline
 $r \leq r_{min}$ & $r_{min} < r < r_{inv}$ & $r_{inv} \leq r < r_{max}$ & $r \geq r_{max}$ \\
 \hline
  A. \eqref{low_nrg} & C (17 $\times$  35) & C (17 $\times$  84) & E. \eqref{high_nrg_rel_w}\\
  A. \eqref{low_nrg} & C (12 $\times$  37) & C (10 $\times$  84) & E. \eqref{high_nrg_rel_w} \\
  A. \eqref{low_nrg} & C (8 $\times$  40) & C (7 $\times$  49) & E. \eqref{high_nrg_rel_w}\\
  A. \eqref{low_nrg} & C (12 $\times$  42) & C (9 $\times$  53) & E. \eqref{high_nrg_rel_w}\\
  A. \eqref{low_nrg} & C (8 $\times$  45) & C (7 $\times$  60) & E. \eqref{high_nrg_rel_w}\\
 \hline
\end{tabular}}
\caption{Summary of the methods employed by the functions \texttt{SFQED\_INV\_COMPTON\_rate} and \texttt{SFQED\_LCFA\_INV\_COMPTON\_PHOTON\_energy} in each region of their domain. ``C'' denotes Clenshaw's recurrence applied to a Chebyshev expansion, with the number inside the round brackets reporting the available number of Chebyshev coefficients. \change{The value of the Chebyshev coefficients is available in the ``coefficients'' folder of SFQEDtoolkit, see, e.g., \url{https://github.com/QuantumPlasma/SFQEDtoolkit}.} ``A.'' denotes the asymptotic approximation in Eq.~\eqref{low_nrg}, while ``E.'' denotes the exponential approximation in Eq.~\eqref{high_nrg_rel_w}.}
\label{summary_phtn}
\end{table}
As briefly discussed, each computational cycle involving SFQED processes first consists in determining whether an event occurs according to its rate. From Eq.~\eqref{pe_dp}, the rate of photon emission is~\cite{Baier-book}
\begin{equation} \label{pe_rate}
R_{pe}(\varepsilon_e, \chi_e) = \int_0^{\varepsilon_e}{\frac{d^2W_{pe}}{dt d\varepsilon_\gamma}(\varepsilon_\gamma, \varepsilon_e, \chi_e) d\varepsilon_\gamma} = \frac{\alpha m_e^2 c^4}{3 \sqrt{3} \pi \hbar \varepsilon_e} \int_{0}^{\infty}{\frac{5u^2 + 7u + 5}{(1+u)^3} \mathrm{K}_{\frac{2}{3}} \biggl(\frac{2u}{3\chi_e}\biggr) du}.
\end{equation}
For its implementation in SFQEDtoolkit, it is convenient to change the dummy variable of integration from $u$ to $v = 2u/3\chi_e$ and express all quantities in normalized units. Namely, we use an angular frequency $\omega_r$ as a reference, and consequently obtain a reference time $T_r=1/\omega_r$, a reference length $\lambda_r=c/\omega_r$, and a reference field $E_r=m_e c \omega_r/|e|$, while electron and photon energies are normalized as  $\gamma_{e/\gamma} = \varepsilon_{e/\gamma} / m_e c^2$. Notice that, e.g., in a simulation where lengths are in units of the laser wavelength $\lambda$, then $\omega_r = 2 \pi c/ \lambda$ and $\lambda_r = \lambda / 2 \pi$. In practice, $\omega_r$ should be an important frequency that we want to resolve. The use of normalized quantities exhibits the scale invariance of the Lorentz equation, and avoids to incur in possible numerical issues related to the use of numbers that are too big or too small for floating-point arithmetic. After the above transformations, Eq.~\eqref{pe_rate} becomes
\begin{equation} \label{pe_rate_3}
W_{\text{rad}}(\gamma_e, \chi_e) = \frac{R_{pe}(\gamma_e, \chi_e)}{\omega_r} = \frac{\alpha}{\sqrt{3}\pi}\frac{\lambda_r}{\lambda_C}\frac{\chi_e}{\gamma_e} \tilde{W}_{\text{rad}}(\chi_e),
\end{equation}
where $\lambda_C = \hbar / m_e c$ is the reduced Compton length and\footnote{In practice, the upper bound of integration is 700 as the integrand becomes so small that the whole integral from 700 to infinity is $\lesssim 10^{-305}$ leading to underflow even with double precision accuracy.}
\begin{equation} \label{wtilde}
\tilde{W}_{\text{rad}}(\chi_e) =  \int_{0}^{\infty}\frac{45 (v \chi_e)^2 + 42 v \chi_e + 20}{(2 + 3 v \chi_e)^3} \mathrm{K}_{\frac{2}{3}} \bigl(v \bigr) dv.
\end{equation}
In order to implement Eq.~\eqref{pe_rate_3}, we only need to compute $\mathcal{C}[\tilde{W}_{\text{rad}}(\chi_e)]$ to the desired accuracy $\Delta_{pe} = \left| \left\{\tilde{W}_{\text{rad}}(\chi_e) - \mathcal{C}[\tilde{W}_{\text{rad}}(\chi_e) \right\} / \tilde{W}_{\text{rad}}(\chi_e) \right|$. Figure~\ref{phtn_rate_acc} displays $\Delta_{pe}$ over the considered interval $0 \leq \chi_e \leq 2000$ clearly showing SFQEDtoolkit's better than 0.1\% accuracy throughout its domain. In SFQEDtoolkit, the Chebyshev coefficients of $\mathcal{C}[\tilde{W}_{\text{rad}}(\chi_e)]$ are precomputed and stored into five separate text files according to the following ranges: $0 \leq \chi_e < 2$, $2 \leq \chi_e < 20$, $20 \leq \chi_e < 80$, $80 \leq \chi_e < 600$, and $600 \leq \chi_e \leq 2000$ (see the summary in Tab.~\ref{summary_phtn}). At runtime, coefficients are loaded once when the simulation is initialized. When the rate is calculated by calling the function \texttt{SFQED\_INV\_COMPTON\_rate}, the relevant interval is selected depending on $\chi_e$, and Clenshaw's recurrence formula is applied to the corresponding coefficients (see \ref{clensh}).
Approximately, ten coefficients per interval are needed to compute $\mathcal{C}[\tilde{W}_{\text{rad}}(\chi_e)]$ with the desired accuracy (see Tab.~\ref{summary_phtn}). Such a small number of coefficients can fit in a modern CPU L1 cache greatly speeding up the simulation. Given the smoothness of $\mathcal{C}[\tilde{W}_{\text{rad}}(\chi_e)]$, no asymptotic expansion is used.
\captionsetup{skip=3pt}

We stress that the fast and accurate calculation of SFQED rates has major implications on the simulation results. On the one hand, insufficient accuracy may result into a significant error in the predicted number of events, which systematically accumulates during the simulation. On the other hand, low performances may noticeably slow down the simulation given that SFQED rates are evaluated at each timestep and for each particle. For example, we implemented SFQEDtoolkit in Smilei and simulated the evolution of an ultralow-density bunch of $N_e = 10^{10}$ electrons \change{with 10~GeV energy} in a constant and uniform magnetic field. Parameters were chosen such that $\chi_e = 1$, initially, and the duration of the simulation was  $T_{sim} = \text{1.3~fs}$, i.e., approximately half of the mean time required for an electron to emit once. By comparing the number of photons produced in the Smilei simulation with SFQEDtoolkit $N_{tk} = 4.834 \times 10^9$ \change{and }in the Smilei simulation with its default 256 points table $N_{S} = 5.111 \times 10^9$, with the expected number of photons $N_{a} = N_e T_{sim} R_{pe}(\varepsilon_e, \chi_e = 1) \approx 4.835 \times 10^9$, we observe that even after such a short amount of time Smilei's 256 points table value $N_{S}$ differs by 5.70\% from the analytical prediction, while SFQEDtoolkit's value $N_{tk}$ differs by 0.04\%. Finally, we recall that the above LCFA rates are suited for both the local probability and the optical depth method (see Sec.~\ref{probabs}).

\subsection{Pair creation rate}  \label{pa_pr_rate}

The treatment of the pair creation rate follows similar steps as those of the photon emission rate. By using the symmetry with respect to the $\varepsilon_\gamma/2$-axis of the integrand in Eq.~\eqref{pp_dp} and by changing the variable of integration to $v$ with $0<v<1$ and $\varepsilon_e = \varepsilon_\gamma (1+v)/2$, the rate of photon conversion into an electron positron pair is~\cite{Baier-book}
\begin{equation} \label{pp_tot}
R_{pp}(\varepsilon_\gamma, \chi_{\gamma}) = \int_0^{\varepsilon_\gamma}{\frac{d^2W_{pp}}{dt d\varepsilon_e}(\varepsilon_e, \varepsilon_\gamma, \chi_\gamma) d\varepsilon_e} = \frac{\alpha m_e^2 c^4}{3 \sqrt{3} \pi \hbar \epsilon_\gamma} \int_{0}^{1}{dv \frac{9-v^2}{1-v^2} \mathrm{K}_{\frac{2}{3}}\biggl(\frac{8}{3 \chi_\gamma (1-v^2)}\biggr)}.
\end{equation}
Notice that in Eq.~\eqref{pp_tot} the ultrarelativistic assumption $\varepsilon_\gamma \gg m_e c^2$ and $\varepsilon_e \gg m_e c^2$ allowed us to approximate the upper and lower limits of integration to $\varepsilon_\gamma$ and zero, respectively. By converting to normalized units, we get
\begin{equation} \label{pp_tot_real}
W_{\text{pair}}(\gamma_\gamma,\chi_{\gamma}) = \frac{R_{pp}(\varepsilon_\gamma, \chi_{\gamma})}{\omega_r} = \frac{\alpha}{\sqrt{3} \pi}\frac{\lambda_r}{\lambda_C}\frac{1}{\gamma_\gamma}\tilde{W}_{\text{pair}}(\chi_\gamma),
\end{equation}
where
\begin{equation} \label{wtilde_pair}
\tilde{W}_{\text{pair}}(\chi_\gamma) = \int_{0}^{1}{dv \frac{9-v^2}{3(1-v^2)} \mathrm{K}_{\frac{2}{3}}\biggl(\frac{8}{3 \chi_\gamma (1-v^2)}\biggr)}.
\end{equation}
Thus, we only need to approximate $\tilde{W}_{\text{pair}}(\chi_\gamma)$ to the desired accuracy $\Delta_{pp}$ defined, as in the photon case, as the relative error between the exact and the approximate function. The considered range $0.01 \leq \chi_\gamma \leq 2000$ was divided into seven intervals: $0.01 \leq \chi_\gamma < 0.24$, $0.24 \leq \chi_\gamma < 0.4$, $0.4 \leq \chi_\gamma < 2$, $2 \leq \chi_\gamma < 20$, $20 \leq \chi_\gamma < 80$, $80 \leq \chi_\gamma < 600$ and $600 \leq \chi_\gamma \leq 2000$. While for each interval in $0.24 \leq \chi_\gamma \leq 2000$ the coefficients of $\mathcal{C}[\tilde{W}_{\text{pair}}(\chi_\gamma)]$ were computed, for $ \chi_\gamma < 0.24$ the pair creation rate is exponentially small and a Chebyshev expansion is no longer suited to accurately calculate $\tilde{W}_{\text{pair}}(\chi_\gamma)$. In this range SFQEDtoolkit uses the asymptotic expansion~\cite{Baier-book}
\begin{equation} \label{low_chi}
W_{\text{pair}}(\gamma_\gamma,\chi_{\gamma}) \approx \frac{3 \sqrt{3} \alpha}{16 \sqrt{2}} \frac{\lambda_r}{\lambda_C} \frac{\chi_\gamma}{\gamma_\gamma} 
 e^{-\frac{8}{3\chi_\gamma}} \Bigl(1 - \frac{11}{64}\chi_\gamma + \frac{7585}{73728}\chi_\gamma^2\Bigr) \quad \text{for $\chi_{\gamma} < 0.24$}.
\end{equation}
Note that the pair creation rate becomes negligibly small $W_{\text{pair}} \lesssim 3.5 \times 10^{-119} \alpha \lambda_r/\lambda_C \gamma_\gamma$ for $\chi_\gamma < 0.01$. Hence, the contribution of the regions where $\chi_\gamma < 0.01$ is neglected in SFQEDtoolkit. 
\begin{table}
\resizebox{\textwidth}{!}{
\begin{tabular}{ |p{2.9cm}||p{1.5cm}|  }
 \hline
 \multicolumn{2}{|c|}{\texttt{SFQED\_BREIT\_WHEELER\_rate}} \\
 \hline
 \hspace{1.3cm}$\chi_\gamma$ & \hspace{0.cm}Method\\
 \hline
 $0.01 \leq \chi_\gamma < 0.24$   & A. \eqref{low_chi} \\
 $0.24 \leq \chi_\gamma < 0.4$   & C (7) \\
 $0.4 \leq \chi_\gamma < 2$   & C (11) \\
 $2 \leq \chi_\gamma < 20$  & C (12) \\
 $20 \leq \chi_\gamma < 80$ & C (7) \\
 $80 \leq \chi_\gamma < 600$   & C (10) \\
 $600 \leq \chi_\gamma \leq 2000$  & C (7) \\
 \hline
\end{tabular}
\begin{tabular}{ |p{2.9cm}||p{2.3cm}|p{3.1cm}|p{2.1cm}|  }
 \hline
 \multicolumn{4}{|c|}{\texttt{SFQED\_BREIT\_WHEELER\_ELECTRON\_energy}} \\
 \hline
  \hspace{1.3cm}$\chi_\gamma$ & $0 \leq |r^{\prime}| < 0.9$ & $0.9 \leq |r^{\prime}| \leq 0.9999$ & $|r^{\prime}| > 0.9999$ \\
 \hline
  $0.01 \leq \chi_\gamma \leq 0.3$ & C (15 $\times$  12) & C (15 $\times$  63) & E. \eqref{inversion}\\  
  $0.3 \leq \chi_\gamma \leq 2$ & C (9 $\times$  11) & C (9 $\times$  54) & E. \eqref{inversion}\\
   $2 \leq \chi_\gamma \leq 20$ & C (11 $\times$  9) & C (12 $\times$  37) & E. \eqref{inversion}\\
   $20 \leq \chi_\gamma \leq 80$ & C (7 $\times$  8) & C (7 $\times$  23) & E. \eqref{inversion}\\
   $80 \leq \chi_\gamma \leq 600$ & C (9 $\times$  9) & C (8 $\times$  14) & E. \eqref{inversion}\\
   $600 \leq \chi_\gamma \leq 2000$ & C (5 $\times$  9) & C (5 $\times$  8) & E. \eqref{inversion}\\
 \hline
\end{tabular}}
\caption{Summary of the methods employed by the functions \texttt{SFQED\_BREIT\_WHEELER\_rate} and \texttt{SFQED\_BREIT\_WHEELER\_ELECTRON\_energy} in each region of the computational domain. ``C'' denotes Clenshaw's recurrence applied to a Chebyshev expansion, with the number inside the round brackets reporting the available number of Chebyshev coefficients. \change{The value of the Chebyshev coefficients is available in the ``coefficients'' folder of SFQEDtoolkit, see, e.g., \url{https://github.com/QuantumPlasma/SFQEDtoolkit}.} ``A.'' denotes the asymptotic approximation in Eq.~\eqref{low_chi}, while ``E.'' denotes the exponential approximation in Eq.~\eqref{inversion}.}
\label{summary_pair}
\end{table}

It is worth noticing that, \change{when using simulations to model realistic scenarios,} the value of $\chi_\gamma$ is necessarily affected by an \change{experimental or observational uncertainty $\Delta \chi_\gamma$, e.g., because of the limited knowledge of physical parameters such as the electromagnetic field or the particle's energy, and the simulation itself is affected by numerical and round-off effects}. This uncertainty $\Delta \chi_\gamma$ propagates to $W_{\text{pair}}$ as $\Delta W_{\text{pair}} \approx (d W_{\text{pair}}/d \chi_\gamma) \Delta \chi_\gamma$. Since the relative error $d \ln(W_{\text{pair}})/d \chi_\gamma$ rapidly diverges for $\chi_\gamma \rightarrow 0$, not only $W_{\text{pair}}$ is tiny for small $\chi_\gamma$, but also its relative error necessarily becomes large. For this reason \change{it is often of limited significance to consider SFQED pair production in regions where  $\chi_\gamma \ll 1$. In} addition to the function \texttt{SFQED\_BREIT\_WHEELER\_rate} which returns $W_{\text{pair}}(\gamma_\gamma,\chi_{\gamma})$ for $0.01 \leq \chi_\gamma \leq 2000$, SFQEDtoolkit also provides a function \texttt{SFQED\_BREIT\_WHEELER\_rate\_fast} which returns zero for $\chi_\gamma < 0.3$. Since $W_{\text{pair}} \lesssim 9.1 \times 10^{-6} \alpha \lambda_r/\lambda_C \gamma_\gamma$ for $\chi_\gamma < 0.3$, the computationally cheaper \texttt{SFQED\_BREIT\_WHEELER\_rate\_fast} is expected to provide essentially the same results as
\texttt{SFQED\_BREIT\_WHEELER\_rate} in many relevant cases while \change{possibly} improving the performance of the simulation.
\captionsetup{skip=3pt}
\begin{figure}[!tb]
\centering
\includegraphics[width=1.\linewidth]{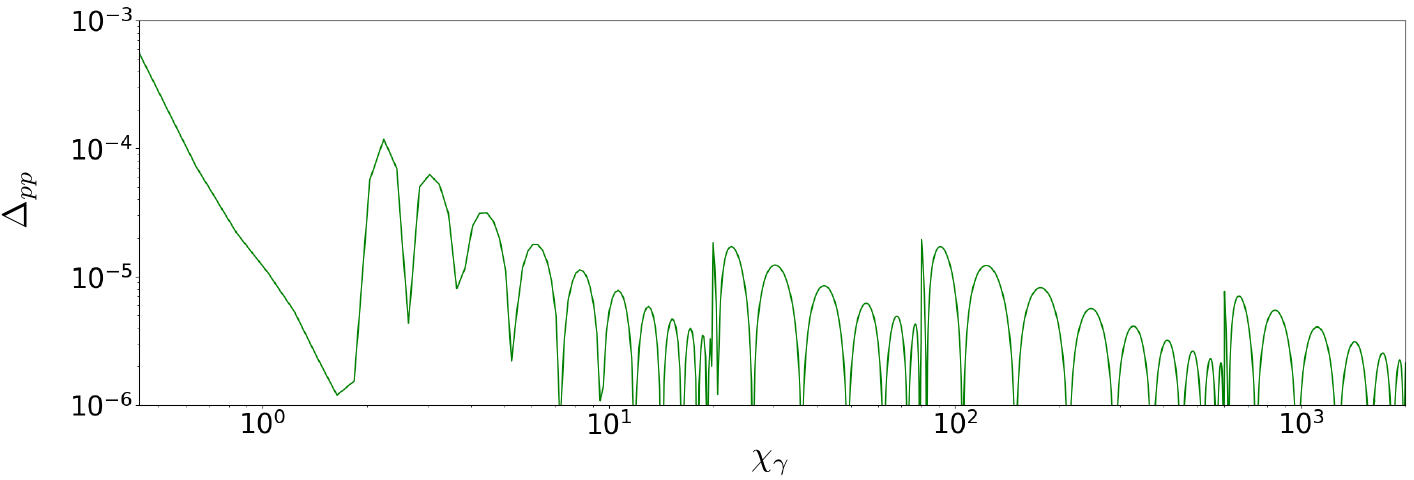}
\caption{Relative accuracy $\Delta_{pp}(\chi_\gamma) = \left| \left\{\tilde{W}_{\text{pair}}(\chi_\gamma) - \mathcal{C}[\tilde{W}_{\text{pair}}(\chi_\gamma) \right\} / \tilde{W}_{\text{pair}}(\chi_\gamma) \right|$ between the analytical and the numerical SFQEDtoolkit photon-to-electron-positron pair conversion rate (see function \texttt{SFQED\_BREIT\_WHEELER\_rate} in \ref{user}). Here $\Delta_{pp}(\chi_\gamma)$ is evaluated at $10^4$ evenly spaced points in the interval $0.01 \leq \chi_e \leq 2000$.}
\label{pair_rate_acc}
\end{figure}

Figure~\ref{pair_rate_acc} displays $\Delta_{pp}$ over the considered domain $0.01 \leq \chi_\gamma \leq 2000$. Similarly to the photon emission rate, approximately ten coefficients per interval are needed to compute $\mathcal{C}[\tilde{W}_{\text{pair}}(\chi_\gamma)]$ with the desired accuracy. Table~\ref{summary_pair} summarizes the strategies of the pair creation routines reported in sections \ref{pa_pr_rate}-\ref{pa_pr}.

\subsection{Photon emission spectrum} \label{ph_em_nrg}

Once an event is deemed to occur, the particles of the final state need to be generated according to the probability distribution of the process. By averaging (summing) over the initial (final) spin and polarization degrees of freedom and by assuming that the momentum of the generated particles is aligned to the momentum of the incoming particle, only the calculation of particles' final energy is needed to determine the state of particles after the event. Now, let us assume that we want to sample a particle with energy $\bar{\varepsilon}$ in the range $0 < \bar{\varepsilon} < \varepsilon_{tot}$ according to an arbitrary distribution $f(x,\varepsilon)$. In this case one can resort to the inverse transform sampling (ITS) method. Namely, given an $f(x, \varepsilon)$, one needs to solve the equation
\begin{equation} \label{ITS_base}
\int_{0}^{\bar{\varepsilon}} f(x, \varepsilon)d\varepsilon - r\int_{0}^{\varepsilon_{tot}}f(x,\varepsilon)d\varepsilon = 0
\end{equation}
in the unknown $\bar{\varepsilon}$. Here $x$ represents generic constant parameters and $0 < r < 1$ is a uniformly distributed random number. For photon emission by an electron, this implies that one needs to solve
\begin{equation} \label{ITS_pe}
\int_{0}^{\bar{\varepsilon}}\frac{d^2W_{pe}}{dtd\varepsilon_\gamma}(\varepsilon_\gamma, \varepsilon_e, \chi_e) d\varepsilon_\gamma - r\int_{0}^{\varepsilon_{e}}\frac{d^2W_{pe}}{dtd\varepsilon_\gamma}(\varepsilon_\gamma, \varepsilon_e, \chi_e)d\varepsilon_\gamma = I_{pe}(\bar{\varepsilon}, \varepsilon_e, \chi_e) - r R_{pe}(\varepsilon_e, \chi_e) = 0,
\end{equation}
where the cumulative distribution function $I_{pe}(\bar{\varepsilon}, \varepsilon_e, \chi_e)$ is obtained by integrating Eq.~\eqref{pe_dp} up to an arbitrary energy value $\bar{\varepsilon}$
\begin{equation} \label{pe_r_prtl}
I_{pe}(\bar{\varepsilon}, \varepsilon_e, \chi_e) = \int_{0}^{\bar{\varepsilon}}\frac{d^2W_{pe}}{dtd\varepsilon_\gamma}(\varepsilon_e, \varepsilon_\gamma, \chi_e)d\varepsilon_\gamma.
\end{equation}
Equation~\eqref{ITS_pe} defines an implicit function $\bar{\varepsilon} = G_{pe}(r, \varepsilon_e, \chi_e)$, which can be calculated up to arbitrary accuracy by numerically solving Eq.~\eqref{ITS_pe} with a suitable root-finding algorithm such as the Brent-Dekker method~\cite{brentCJ71, dekker69}. However, direct application of the above recipe at runtime in a PIC code would be prohibitively expensive. Modified Bessel functions of the second kind $\mathrm{K}_{2/3}(x)$ and $\int_{\eta}^{\infty}{dy \mathrm{K}_{1/3}(y)}$ as well as several computational cycles where multiple integrals need to be calculated are required to obtain $\bar{\varepsilon}$ from Eq.~\eqref{ITS_pe} via a root-finding routine. 

In SFQEDtoolkit, the function \texttt{SFQED\_LCFA\_INV\_COMPTON\_PHOTON\_energy} provides users with a fast and accurate approximation of $G_{pe}(r, \varepsilon_e, \chi_e)$. The accuracy of the approximation $\Delta_r$ is given by $\Delta_r = \left| \bar{\varepsilon}_{ITS} - \bar{\varepsilon}_{tk} \right| / \bar{\varepsilon}_{ITS}$, where $\bar{\varepsilon}_{ITS}$ is the value obtained via the ITS method computed with more than ten significant digits accuracy, and $\bar{\varepsilon}_{tk}$ is the value returned by SFQEDtoolkit. It is worth noting that \change{although the functions $I_{pe}$ and $R_{pe}$ in Eqs.~\eqref{ITS_pe}-\eqref{pe_r_prtl} depend on two- and three-variables, respectively, the total energy of the parent particle $\varepsilon_{e}$ can be easily factored out, and one is required to approximate functions of only one and two variables, in practice.}

\begin{table}
\hspace{4cm}
\begin{tabular}{ |p{2.8cm}|p{0.9cm}|p{0.9cm}|p{1.2cm}|  }
 \hline
 \hspace{1.3cm}$\chi_e$ & $r_{min}$ & $r_{inv}$ & $r_{max}$ \\
 \hline
  $0 \leq \chi_e < 2$ & 0.028 & 0.986 & 0.99999\\
  $2 \leq \chi_e < 20$ & 0.041 & 0.987 & 0.99999\\
  $20 \leq \chi_e < 80$ & 0.045 & 0.987 & 0.99996\\
  $80 \leq \chi_e < 600$ & 0.040 & 0.987 & 0.99997\\
  $600 \leq \chi_e \leq 2000$ & 0.050 & 0.987 & 0.99998\\
 \hline
\end{tabular}
\caption{The numerical values of $r_{min}$, $r_{inv}$ and $r_{max}$ for each of the five intervals of $\chi_e$.}
\label{summary_r}
\end{table}

\begin{figure}[!tb]
\centering
\includegraphics[width=1.\linewidth]{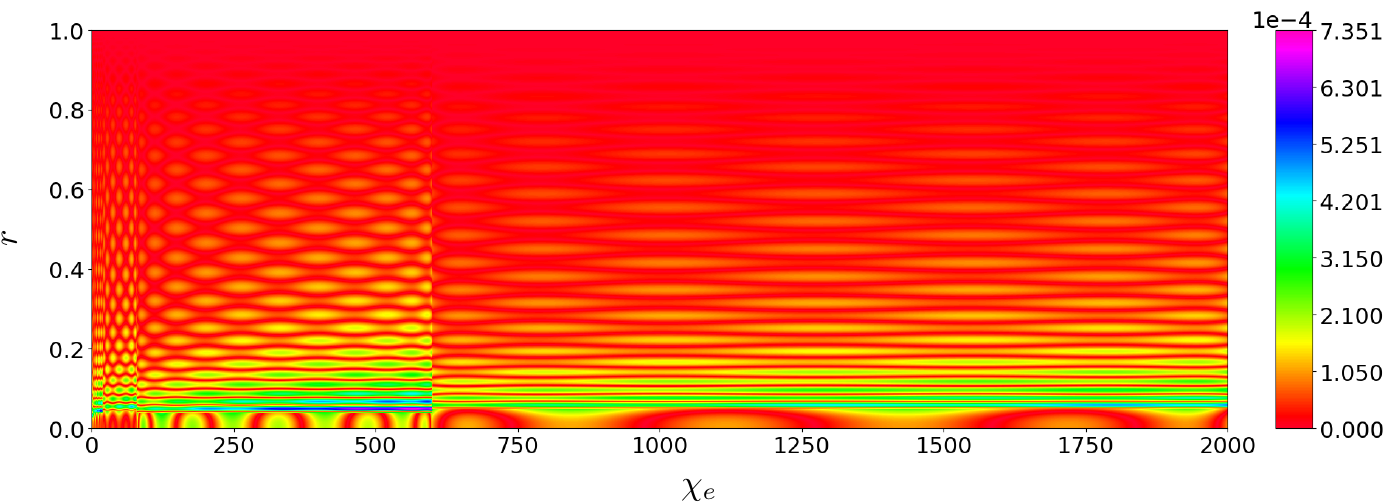}
\caption{Relative difference $\Delta_r(\chi_e, r)$ between the exact and the SFQEDtoolkit computed photon emission energies (see \texttt{SFQED\_LCFA\_INV\_COMPTON\_PHOTON\_energy} in \ref{user}). The contour plot is obtained by evenly evaluating $\Delta_r(\chi_e,r)$ at $10^4(\chi_e)\times 10^3(r)$ points in the domain $0 \leq \chi_e \leq 2000$, $0 \leq r \leq 1$. The colorbar ranges from the lowest to the highest recorded $\Delta_r$.}
\label{phtn_acc}
\end{figure}

Similarly to the photon emission rate, the domain of $\chi_e$ and $r$ was divided into smaller intervals either to reduce the required number of Chebyshev coefficients or to use a different approximation method. The same intervals as for the photon emission rate were used for $\chi_e$, while $r$ was divided into four intervals: $0 < r \leq r_{min}$, $r_{min} < r \leq r_{inv}$, $r_{inv} < r \leq r_{max}$, and $r_{max} < r < 1$. The value of $r_{min}$, $r_{inv}$ and $r_{max}$ depends on $\chi_e$, and is reported in Tab.~\ref{summary_r}. Table~\ref{summary_phtn} summarizes the decomposition of the domain, the method used in each interval and, when the function is approximated with Chebyshev polynomials, the number of employed Chebyshev coefficients. Figure~\ref{phtn_acc} plots the value of $\Delta_r$ for the whole range of $0 \leq \chi_e \leq 2000$ and $0 < r < 1$, which shows that $\Delta_r < 10^{-4}$ throughout the whole considered domain.
\captionsetup{skip=3pt}

For its implementation in SFQEDtoolkit, the function $I_{pe}$ in Eq.~\eqref{pe_r_prtl} was converted to normalized units, and the dummy variable of integration was changed from $\varepsilon_\gamma$ to 
\begin{equation} \label{subs_altri}
w = \sqrt[3]{\frac{2 \varepsilon_\gamma}{3(\varepsilon_e - \varepsilon_\gamma)\chi_e}};
\qquad
\varepsilon_\gamma = \frac{3 \varepsilon_e \chi_e w^3}{2 + 3 \chi_e w^3},
\end{equation}
which gives
\begin{align} \label{pe_eff_int}
I_{pe}(&\bar{w},\varepsilon_e,\chi_e) =  \frac{\alpha}{\sqrt{3}\pi}\frac{\lambda_r}{\lambda_C}\frac{\chi_e}{\gamma_e}\omega_r\tilde{I}_{pe}(\bar{w},\chi_e),
\end{align}
where
\begin{align} \label{pe_eff_int_2}
\tilde{I}_{pe}(\bar{w}, \chi_e) = & \int_{0}^{\bar{w}}\frac{9 w^2 dw}{2 (1 + s)^3}
\biggl\{\left[1 + (1 + s)^2 \right] \mathrm{K}_{\frac{2}{3}}(w^3) - (1 + s) \int_{w^3}^{\infty} \mathrm{K}_{\frac{1}{3}}(y)dy \biggr\},
\end{align}
and  $s=3\chi_e w^3/2$. By comparing Eqs.~\eqref{pe_rate_3} and \eqref{pe_eff_int}, one finds that the ITS equation \eqref{ITS_pe} simplifies to
\begin{equation} \label{again}
\tilde{I}_{pe}(\bar{w}, \chi_e) - r \tilde{W}_{\text{rad}}(\chi_e) = 0.
\end{equation}
From $\bar{w}$, the photon energy $\varepsilon_\gamma$ is easily obtained via Eq.~\eqref{subs_altri}. The change of variable defined in Eq.~\eqref{subs_altri} improves the smoothness of $I_{pe}$ by removing the $\varepsilon_\gamma^{-2/3}$ singularity of the integrand for $\varepsilon_\gamma \rightarrow 0$, thereby reducing the required number of Chebyshev coefficients. 

We begin describing how SFQEDtoolkit implements the two-variable function $\bar{w} = \tilde{G}_{pe}(r, \chi_e)$ implicitly defined by Eq.~\eqref{again} in the interval $r_{min} < r < r_{max}$. A natural choice is to approximate $\tilde{G}_{pe}(r, \chi_e)$ with Chebyshev polynomials. In practice, despite this approach performs extremely well almost everywhere, $\tilde{G}_{pe}(r, \chi_e)$ and its derivatives rapidly grow for $r \to r_{max}$ such that the number of Chebyshev coefficients needed to accurately approximate the region around $r_{max}$ becomes large. To reduce the number of required coefficients while retaining high accuracy, in the interval $r_{min} < r < r_{inv}$ and $r_{inv} < r < r_{max}$ the Chebyshev expansion of $\tilde{G}_{pe}(r, \chi_e)$ and of $1/\tilde{G}_{pe}(r, \chi_e)$ is employed, respectively. In the latter case, $\tilde{G}_{pe}(r, \chi_e)$ is then readily obtained from $[1/\tilde{G}_{pe}(r, \chi_e)]^{-1}$. 

\change{Finally}, SFQEDtoolkit solves Eq.~\eqref{again} in the limit $r \to 0$ and $r \to 1$, i.e., when the lower and the higher energy tail of the photon spectrum are approached, by resorting to asymptotic expansions and exponential approximations. For $r \to 0$, one can expand Eq.~\eqref{pe_eff_int_2} as
\begin{equation} \label{pe_eff_int_approx_w_small}
\tilde{I}_{pe}(\bar{w},\chi_e)
\xrightarrow{\bar{w} \to 0}
\frac{9}{2^{1/3}}\Gamma\biggl(\frac{2}{3}\biggr)\bar{w},
\end{equation}
which is a better than 0.1\% approximation of $\tilde{I}_{pe}(\bar{w},\chi_e)$ for $r \leq r_{min}$. By substituting Eq.~\eqref{pe_eff_int_approx_w_small} in Eq.~\eqref{again}, one immediately obtains
\begin{equation}
\bar{w} = r \tilde{W}_{\text{rad}}(\chi_e)\biggl[\frac{9}{2^{1/3}}\Gamma\biggl(\frac{2}{3}\biggr)\biggr]^{-1}.
\label{low_nrg}
\end{equation}
Regarding $r \to 1$, i.e., $\bar{w} \to \infty$, we leverage on the fact that for $\bar{w}$ above a threshold $w_0$ the function $\tilde{I}_{pe}(\bar{w},\chi_e)$ saturates to $\tilde{W}_{\text{rad}}(\chi_e) = \tilde{I}_{pe}(\infty,\chi_e)$, and the following exponential approximation holds
\begin{equation} \label{high_nrg_rel}
\tilde{I}_{pe}(\bar{w}, \chi_e) \approx \tilde{W}_{\text{rad}}(\chi_e) (1 - e^{-(\bar{w}^3-w_0^3)}) +  \tilde{I}_{pe}(w_0, \chi_e) e^{-(\bar{w}^3-w_0^3)}.
\end{equation}
By substituting Eq.~\eqref{high_nrg_rel} in Eq.~\eqref{again} one obtains
\begin{equation} \label{high_nrg_rel_w}
\bar{w} = \sqrt[3]{w_0^3 - \log \biggl[ \frac{\mathcal{C}[\tilde{W}_{\text{rad}}(\chi_e)] (1-r)}{\mathcal{C}[\tilde{W}_{\text{rad}}(\chi_e)] - \mathcal{C}[\tilde{I}_{pe}(w_0, \chi_e)]}\biggr]}.
\end{equation}
Note that: (i) there is no need to calculate the cubic root in Eq.~\eqref{high_nrg_rel_w} since $\varepsilon_\gamma$ depends \change{only }on $\bar{w}^3$; (ii) in Eq.~\eqref{high_nrg_rel_w} we have explicitly indicated that in SFQEDtoolkit the Chebyshev expansion of $\tilde{W}_{\text{rad}}(\chi_e)$ and $\tilde{I}_{pe}(w_0, \chi_e)$ \change{are} used; (iii) this approximation is used for $r \geq r_{max}$, and the corresponding $\chi_e$-dependent $w_0$ is obtained from $w_0 = \mathcal{C}[\tilde{G}_{pe}(r_{max}, \chi_e)]$ where $\mathcal{C}[\tilde{G}_{pe}(r_{max}, \chi_e)]$ is the Chebyshev approximation of the function obtained by setting $r = r_{max}$ in Eq.~\eqref{again}.

\begin{figure}[!tb]
\centering
\includegraphics[width=1.\linewidth]{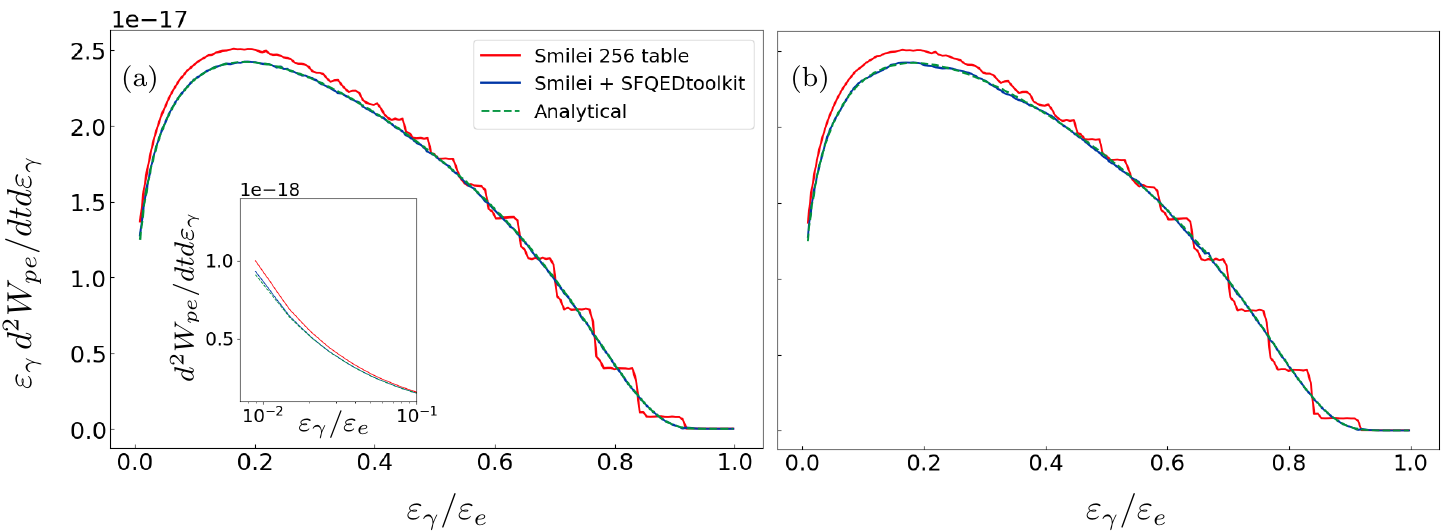}
\caption{Synchrotron emission spectrum (main panel) and the low-energy region of the differential distribution (inset) obtained from the evolution of $10^{10}$ electrons \change{with 10~GeV energy located} in a constant and uniform magnetic field such that $\chi_e = 1$. The results of the simulation performed by using the default $256\times256$ points table of Smilei (red solid line), the results of the same simulation with the SFQEDtoolkit implementation (blue solid line), and the analytical spectrum (green dashed line). The solid blue line in panels (a) and (b) displays the results obtained from the SFQEDtoolkit implementation where the provided $17 \times 35$ Chebyshev coefficient's matrix and the reduced $15 \times 25$ sub-matrix
are used, respectively. See Sec.~\ref{performance} for details.}
\label{smilei_comp}
\end{figure}

We implemented SFQEDtoolkit in the open source PIC code Smilei~\cite{derouillatCPC18} version 4.7 and simulated the evolution of an ensemble of $10^{10}$ electrons. The charge density was kept very low to suppress self-fields, and particles were placed in a constant and uniform external magnetic field \change{with 10~GeV energy}. Simulation parameters were chosen such that $\chi_e = 1$, initially. The total simulation time was set to one tenth of the time expected for an electron to emit once. We then repeated the same simulation using the original version of Smilei, which makes use of 256-points pre-computed lookup tables and compared the results with those obtained with SFQEDtoolkit. Figure~\ref{smilei_comp}(a) displays the photon spectrum obtained from the original version of Smilei (red line), with SFQEDtoolkit (blue line), and the analytical prediction (dashed green line). \change{While Smilei with its default 256-points tables shows a marked stairlike pattern at high photon energies and a significant deviation from the analytic spectrum also for photon energies around $0.2\,\varepsilon_e$, SFQEDtoolkit nearly perfectly matches the analytic spectrum throughout the whole domain. Note that such discrepancies of Smilei with its 256-points tables with respect to the analytic spectrum originate from the coarse tabulation. By employing Smile with 1024-points tables the agreement with the analytic spectrum significantly improves with respect to the 256-points tables, particularly in the region $\varepsilon_\gamma / \varepsilon_e \sim 0.2$, while the stairlike pattern is much milder but remains visible also with 1024-points tables.} Regarding performance, a direct comparison of the execution time of the default version of Smilei and of the SFQEDtoolkit implementation does not directly reflect the improvement, because the total simulation time is strongly determined by the other operations of the PIC loop. Nevertheless, in our tests the implementation with SFQEDtoolkit outperformed the 256 points default Smilei by up to 12\%.
\captionsetup{skip=3pt}

\subsection{Pair creation spectrum} \label{pa_pr}

The same methodology employed to sample the energy of emitted photons applies in the case of photon conversion into an electron-positron pair. One solve\change{s} the equation
\begin{equation} \label{ITS_pp}
\int_{0}^{\bar{\varepsilon}}
\frac{d^2W_{pp}}{dtd\varepsilon_e}(\varepsilon_e, \varepsilon_\gamma, \chi_\gamma) d\varepsilon_e - r \int_{0}^{\varepsilon_\gamma} \frac{d^2W_{pp}}{dtd\varepsilon_e}(\varepsilon_e, \varepsilon_\gamma, \chi_\gamma) d\varepsilon_e = I_{pp}(\bar{\varepsilon}, \varepsilon_\gamma, \chi_\gamma) - r R_{pp}(\varepsilon_\gamma, \chi_\gamma) = 0,
\end{equation}
where $I_{pp}(\bar{\varepsilon}, \varepsilon_\gamma, \chi_\gamma)$ is obtained by integrating Eq.~\eqref{pp_dp} up to $\bar{\varepsilon}$
\begin{equation} \label{pp_cumul}
I_{pp}(\bar{\varepsilon}, \varepsilon_\gamma, \chi_\gamma) = \int_{0}^{\bar{\varepsilon}} \frac{d^2W_{pp}}{dtd\varepsilon_e}(\varepsilon_e, \varepsilon_\gamma, \chi_\gamma) d\varepsilon_e .
\end{equation}
In SFQEDtoolkit, we exploited the symmetry of Eq.~\eqref{pp_cumul} with respect to $\varepsilon_\gamma/2$ to halve the domain of integration to $\varepsilon_\gamma/2 < \varepsilon_e < \varepsilon_\gamma$, converted to normalized units, and changed the dummy variable of integration to
\begin{equation} \label{subs_pair}
v = \frac{2 \varepsilon_e - \varepsilon_\gamma}{\varepsilon_\gamma};
\qquad
\varepsilon_e = \frac{\varepsilon_\gamma (1+v)}{2}.
\end{equation}
Thus, the equation that needs to be solved simplifies to
\begin{equation} \label{again_pp}
\tilde{I}_{pp}(\bar{v}, \chi_\gamma) - |r'| \tilde{W}_{\text{pair}}(\chi_\gamma) = 0
\end{equation}
where
\begin{align} \label{tilde_I_pp}
\tilde{I}_{pp}(\bar{v}, \chi_\gamma) & = \int_{0}^{\bar{v}} \biggl[\frac{2(1 + v^2)}{1 - v^2} \mathrm{K}_{\frac{2}{3}} \left(\eta_v \right) + \int_{\eta_v}^{\infty} \mathrm{K}_{\frac{1}{3}}\left(y\right)dy\biggr]dv,
\end{align}
$\eta_v = 8 /[3 \chi_\gamma (1 - v^2)]$, and $r'=2r-1$ is a random number uniformly distributed in $-1 < r' < 1$ obtained from a random number $r$ defined in $0 < r < 1$. Equation~\eqref{again_pp} defines an implicit function $\bar{v} = \tilde{G}_{pp}(|r'|, \chi_\gamma)$. Once $\bar{v}$ has been determined, the relation
\begin{equation} \label{return_to_nrg}
\varepsilon_e = \frac{\varepsilon_\gamma [1 + \text{sgn}(r') v]}{2}.
\end{equation}
allows us to obtain the electron $\varepsilon_e$ and positron $\varepsilon_p = \varepsilon_\gamma - \varepsilon_e$ energy in their full $(0,\varepsilon_\gamma)$ domain. For $\chi_\gamma$ the domain of the function $\tilde{G}_{pp}(|r'|, \chi_\gamma)$ is divided into six intervals (see Tab.~\ref{summary_pair}), while for $|r'|$ it is divided into three intervals: $0 \leq |r^{\prime}| < 0.9$, $0.9 \leq |r^{\prime}| \leq 0.9999$, and $|r^{\prime}| > 0.9999$. In $0 \leq |r^{\prime}| \leq 0.9999$, Chebyshev polynomials are used to approximate $\tilde{G}_{pp}(|r'|, \chi_\gamma)$ to the desired accuracy. The division into two intervals, i.e., in two distinct set of Chebyshev coefficients, is made only to reduce the number of required coefficients in the more probable region $0 \leq |r^{\prime}| < 0.9$. In fact, when $|r^{\prime}| \to 1$ the number of coefficients necessary to accurately approximate $\tilde{G}_{pp}(|r'|, \chi_\gamma)$ increases dramatically, and to preserve the computational speed granted by working with a restricted set of coefficients, we have chosen to separate the smaller interval requiring a larger number of coefficients $|r^{\prime}| > 0.9$ from the more probable interval $0 \leq |r^{\prime}| < 0.9$. Finally, for $|r^{\prime}| > 0.9999$ an exponential approximation similar to Eq.~\eqref{high_nrg_rel} is used to approximate the cumulative function
\begin{equation} \label{high_nrg_rel_pair}
\tilde{I}_{pp}(\bar{v}, \chi_\gamma) \approx \tilde{W}_{\text{pair}}(\chi_\gamma) \left[1- e^{-\left(\frac{8}{3 \chi_\gamma (1 - v^2)} - \frac{8}{3 \chi_\gamma (1 - v_0^2)}\right)}\right] + \tilde{I}_{pp}(\bar{v}_0, \chi_\gamma) e^{-\left(\frac{8}{3 \chi_\gamma (1 - v^2)} - \frac{8}{3 \chi_\gamma (1 - v_0^2)}\right)}.
\end{equation}
By substituting Eq.~\eqref{high_nrg_rel_pair} in Eq.~\eqref{again_pp} one obtains
\begin{align} \label{inversion}
\bar{v} & = \sqrt{1 - \biggl(\frac{1}{1 - v_0^2} - \frac{3 \chi_\gamma}{8} \log{\biggl[\frac{\mathcal{C}[\tilde{W}_{\text{pair}}(\chi_\gamma)] (1 - r)}{\mathcal{C}[\tilde{W}_{\text{pair}}(\chi_\gamma)] - \mathcal{C}[\tilde{I}_{pp}(\bar{v}_0, \chi_\gamma)]} \biggr]} \biggr)^{-1}},
\end{align}
where we have indicated that in SFQEDtoolkit a Chebyshev expansion is used to compute $\tilde{W}_{\text{pair}}(\chi_\gamma)$ as well as $\tilde{I}_{pp}(\bar{v}_0, \chi_\gamma)$, and $v_0 = \mathcal{C}[\tilde{G}_{pp}(0.9999, \chi_\gamma)]$. 

\begin{figure}[!tb]
\centering
\includegraphics[width=1.\linewidth]{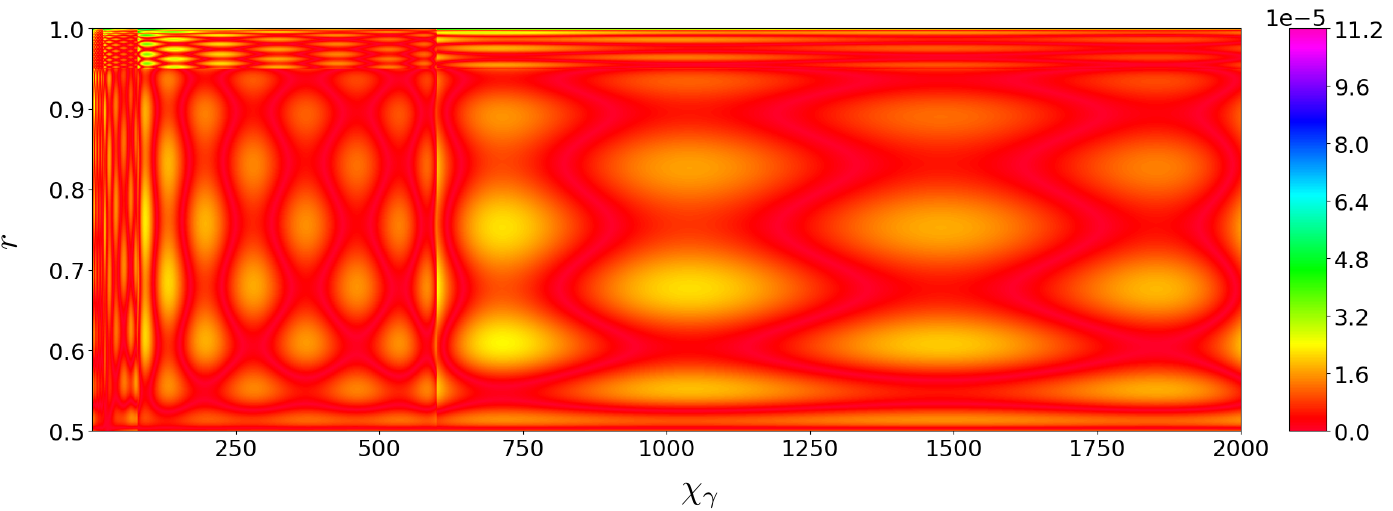}
\caption{Relative difference $\Delta_r(\chi_\gamma, r)$ between the exact and the SFQEDtoolkit computed lepton energy in the nonlinear Breit-Wheeler conversion of an energetic photon in an electron-positron pair (see \texttt{SFQED\_BREIT\_WHEELER\_ELECTRON\_energy} in \ref{user}). The contour plot is obtained by evenly evaluating $\Delta_r(\chi_\gamma, r)$ at $10^4(\chi_\gamma)\times 10^3(r)$ points in the domain $0.3 \leq \chi_e \leq 2000$, $0 \leq r \leq 1$. The colorbar ranges from the lowest to the highest recorded $\Delta_r$.}
\label{pair_acc}
\end{figure}

The function \texttt{SFQED\_BREIT\_WHEELER\_ELECTRON\_energy} returns the energy of the created electron once the normalized photon energy $\gamma_\gamma$, quantum parameter $\chi_\gamma$ and a random number $r$ uniformly distributed in $(0,1)$ are provided as input parameters. Figure~\ref{pair_acc} displays the accuracy of the approximation $\Delta_r = \left| \bar{\varepsilon}_{ITS} - \bar{\varepsilon}_{tk} \right| / \bar{\varepsilon}_{ITS}$ in the domain $0.3 < \chi_\gamma < 2000$ and $0< r <1$. Symbols have the same meaning as in the photon emission case.

\captionsetup{skip=3pt}

\subsection{Changing the number of coefficients used in a simulation} \label{performance}

SFQEDtoolkit allows users to reduce the number of Chebyshev coefficients to be employed at runtime therefore reducing the accuracy but possibly enhancing the performance of the library. For instance, the Chebyshev coefficients $c_i$ of a one-variable function $f(x)$ are stored in text files where the first row is an integer reporting the total number of coefficients $n$ stored in the file, the second and third row contain floating point numbers that report the minimum and the maximum of the interval where the function is approximated with Chebyshev polynomials, respectively, while from the fourth row on the coefficients $c_i$ are written row by row. By adding a colon followed by an integer $k<n$, i.e., by replacing $n$ with $n:k$, only the leading $k$ coefficients out of the $n$ total are employed at runtime. 

Similarly, the Chebyshev coefficients $c_{ij}$ of a two-variable function $f(x,y)$ are stored in text files where the first and fourth row report the number of columns $n$ and of rows $m$ of $c_{ij}$, respectively. The second and third (fifth and sixth) row of the file report the minimum and the maximum of first $x$ (second $y$) variable of $f(x,y)$, respectively. Finally, starting from the seventh row of the file the coefficients $c_{ij}$ are written row by row following a row-major order. Also in this case, one can reduce the number of used coefficients by editing the first (fourth) row of the file, i.e, replacing $n$ ($m$) with $n:k$ ($m:l$) where $k$ ($l$) is an integer smaller than $n$ ($m$). Unless otherwise specified, SFQEDtoolkit uses all the coefficients available in a file.

For example, by repeating the simulation reported in Fig.~\ref{smilei_comp}(a) with a set of $15 \times 25$ coefficients instead of the $17 \times 35$ provided with SFQEDtoolkit, the relative error rises up to 0.7\% while the time to execute the specific task of calculating the photon spectrum reduces by approximately 30\%. Figure~\ref{smilei_comp}(b) displays the photon spectrum obtained with the above-mentioned reduced set of coefficients. Even with the reduced set of coefficients, SFQEDtoolkit provides a photon spectrum in manifestly much better agreement with the analytical prediction than Smilei's default algorithm and $256 \times 256$ points table [see Fig.~\ref{smilei_comp}(b)].

\section{Photon emission beyond the locally-constant-field approximation} \label{blcfa}

In the following we detail how the method of photon emission beyond LCFA developed in Ref.~\cite{dipiazzaPRA19} is implemented in SFQEDtoolkit. In order to retain flexibility and allow users to better adapt and customize functionality to their codes, SFQEDtoolkit provides a set of independent functions each carrying out a specific task of the algorithm.

We begin by briefly reviewing the rationale of the algorithm and explaining how it benefits from a local probability approach instead of the optical depth method (see Sec.~\ref{probabs}). As detailed in Ref.~\cite{dipiazzaPRA19}, the differential probability of photon emission $d^2W_{pe}/dtd\gamma_\gamma$ becomes almost flat for a broad range of photon energies below a threshold $\gamma_{\gamma,\text{LCFA}}$, here normalized to $m_e c^2$, while it nearly coincides with the LCFA distribution above this threshold. This implies that the use of the LCFA rate systematically results in an \change{orders of magnitude} overestimated number of emitted photons \change{for $\varepsilon_\gamma \lesssim (\chi_e/\xi^3) \varepsilon_e$}. Notice that $\gamma_{\gamma,\text{LCFA}}$ depends on a characteristic local time of variation $\tau$ of the transverse Lorentz force $\bm{F}_{L,\perp}$ experienced by the emitting particle. This characteristic time $\tau$ is obtained from the first $\dot{\bm{F}}_{L,\perp}$ and second $\ddot{\bm{F}}_{L,\perp}$ time derivative of $\bm{F}_{L,\perp}$ calculated along the emitting particle's trajectory~\cite{dipiazzaPRA19}. 

\change{We stress} that the calculation of the improved rate of photon emission for each particle and at each timestep, which is required by the optical depth method, now becomes relatively complex and computationally expensive. A simpler and more efficient option is to use the LCFA rate of photon emission to determine whether a photon emission event occurs, locally. Only if the event is deemed to occur according to the LCFA model, then the expected photon energy is sampled from the LCFA differential probability of photon emission. If the sampled photon energy $\bar{\gamma}_\gamma$ exceeds the threshold $\gamma_{\gamma,\text{LCFA}}$, a photon with energy $\bar{\gamma}_\gamma$ is generated and the emitting particle recoils. Otherwise, the photon emission event is either accepted or rejected by comparing an independent uniformly distributed random number in $(0,1)$ with the ratio between the LCFA and the improved differential probability of photon emission calculated at $\bar{\gamma}_\gamma$ (see below). This greatly improves the performance of the code given that most of the above-mentioned extra computational steps are performed rarely and only when needed.

SFQEDtoolkit provides a \texttt{C++} object named \texttt{BLCFA\_Object} which contains (i) a three-element double precision array to store the transverse Lorentz force at the penultimate timestep, (ii) a three-element double precision array to store the difference between the Lorentz force at the penultimate and the antipenultimate timestep, and (iii) a boolean signaling whether the particle was created at the penultimate timestep. \change{This boolean is needed because for a new particle the transverse Lorentz force at the penultimate and antipenultimate timestep is not available, and either the LCFA is employed or no emission occurs for the first timestep after the particle is created (see Ref.~\cite{dipiazzaPRA19} and below).} These quantities together with the simulation timestep $\Delta t$ and the emitting particle energy $\gamma_e$ and quantum parameter $\chi_e$ are used by the routine \texttt{SFQED\_BLCFA\_INV\_COMPTON\_PHOTON\_threshold} to calculate the threshold $\gamma_{\gamma,\text{LCFA}}$ (see below and \ref{user}). In addition, the class \texttt{BLCFA\_Object} serves to derive through inheritance \change{an extended} \texttt{C++} object, which can be used to include all needed information on the state of a computational particle. In the given PIC or Monte Carlo code where the user wants to implement SFQEDtoolkit, a linked list, or possibly better for memory locality, an array or a \texttt{C++} vector class of particle objects derived from \texttt{BLCFA\_Object} can be used to describe the state of an arbitrary number of computational particles. Each \texttt{BLCFA\_Object} is created by calling the routine \texttt{SFQED\_CREATE\_BLCFA\_OBJECT}, and its content is updated at each timestep and for each particle via the routine \texttt{SFQED\_BLCFA\_OBJECT\_update} (see \ref{user} for details). 

For clarity and completeness, we summarize below the steps needed to calculate $\gamma_{\gamma,\text{LCFA}}^{(n)}$ at timestep $(n)$ as well as the acceptance-rejection method. Following Ref.~\cite{dipiazzaPRA19}, by starting from $\bm{x}^{(n)}$ and $\bm{p}^{(n-1/2)}$ advance the momentum to $\bm{p}_{L}^{(n+1/2)}$ with the Lorentz force integrator existing in the code\footnote{Here we consider a leapfrog integrator, i.e., position and momentum are shifted by half step. There is however no conceptual difference in the method if position and momentum are given at the same timestep, such as in a Runge-Kutta method.}. If the particle was created at the penultimate timestep, \change{which is signaled, e.g., by the boolean of the \texttt{BLCFA\_Object},} set $\bm{F}_{L,\perp}^{(n-2)} = \bm{F}_{L,\perp}^{(n-1)} = \bm{F}_{L,\perp}^{(n)}$, then update the boolean describing its status, and continue to the next particle. Otherwise, calculate
\begin{align}
\label{fl}
\bm{F}_{L}^{(n)} & = \frac{\bm{p}_{L}^{(n+1/2)}-\bm{p}^{(n-1/2)}}{\Delta t}, \\
\bm{p}_{L}^{(n)} & = \frac{\bm{p}_{L}^{(n+1/2)}+\bm{p}^{(n-1/2)}}{2}, \\
\label{gammaBLCFA}
\gamma^{(n)}_e & = \sqrt{1+[\bm{p}_{L}^{(n)}]^2}, \\
\label{fl_perp}
\bm{F}_{L,\perp}^{(n)} & = \bm{F}_{L}^{(n)} - \frac{\bm{F}_{L}^{(n)} \cdot \bm{p}_{L}^{(n)}}{[\gamma^{(n)}]^2} \bm{p}_{L}^{(n)}, \\
\label{chi_p}
\chi^{(n)}_e & = \tau_C \gamma^{(n)} \sqrt{\left[\bm{F}_{L,\perp}^{(n)}\right]^2},
\end{align}
where $\tau_C$ is the Compton time and normalized units are employed. Note that the ultrarelativistic approximation $\bm{p}_{L}^{(n)} / |\bm{p}_{L}^{(n)}| \approx \bm{p}_{L}^{(n)} / \gamma^{(n)}$ is used in Eq.~\eqref{fl_perp}. Compute 
\begin{align}
\dot{\bm{F}}_{L,\perp}^{(n)} & = \frac{\bm{F}_{L,\perp}^{(n)} - \bm{F}_{L,\perp}^{(n-1)}}{\Delta t}, \\
\ddot{\bm{F}}_{L,\perp}^{(n)} & = \frac{(\bm{F}_{L,\perp}^{(n)} - \bm{F}_{L,\perp}^{(n-1)}) - (\bm{F}_{L,\perp}^{(n-1)} - \bm{F}_{L,\perp}^{(n-2)})}{(\Delta t)^2}, \\
\delta^{(n)} = & \tau_C^2 \left[ (\dot{\bm{F}}_{L,\perp}^{(n)})^2 + |\bm{F}_{L,\perp}^{(n)} \cdot \ddot{\bm{F}}_{L,\perp}^{(n)}| \right]. \label{condit}
\end{align}
If $(\gamma^{(n)})^2 \delta^{(n)}/\zeta^2 > (\chi^{(n)})^2 (\bm{F}_{L,\perp}^{(n)})^2$ and $\chi^{(n)}>\zeta$, where $\zeta$ is a nearly negligible number relative to unity\footnote{Assuming double precision arithmetic $\zeta \approx 2.22 \times 10^{-16}$.}, then calculate $\tau^{(n)}/\tau_C = 2 \pi \sqrt{[\bm{F}_{L,\perp}^{(n)}]^2/\delta^{(n)}}$. Otherwise, the background fields are either basically constant and the LCFA applies throughout the photon spectrum or the quantum parameter $\chi^{(n)}$ is negligibly small. This condition is introduced to avoid numerical issues for constant background fields, where the LCFA holds and $\tau^{(n)}/\tau_C$ diverges. Finally, following Ref.~\cite{dipiazzaPRA19} 
\begin{equation} \label{gammaLCFA}
\gamma_{\gamma,\text{LCFA}}^{(n)} = \frac{0.7 \gamma_e^{(n)}}{1 + \frac{4}{3 \pi \chi_e^{(n)}}\sinh\left[3\sinh^{-1}\left(\frac{\chi_e^{(n)}}{8 \gamma_e^{(n)}}\frac{\tau^{(n)}}{\tau_C}\right)\right]},
\end{equation}
and the condition $\gamma_{\gamma,\text{LCFA}}^{(n)} < 0.75 \gamma_e^{(n)}$ is evaluated. If this latter condition is violated, most of the spectrum is not approximated by the LCFA model, and the formation of the emission probability becomes an intrinsically nonlocal process. In this case we assume that no emission is deemed for the considered particle at this timestep.

Once $\gamma_{\gamma,\text{LCFA}}^{(n)}$ is known, the routine \texttt{SFQED\_BLCFA\_INV\_COMPTON\_PHOTON\_energy} samples the photon energy according the beyond LCFA differential probability of photon emission. Its input arguments are $\gamma^{(n)}_e$, $\chi^{(n)}_e$, $\gamma_{\gamma,\text{LCFA}}^{(n)}$ as well as two independent uniformly distributed random numbers $0< r_1 < 1$ and $0< r_2 < 1$. This routine first calls
\texttt{SFQED\_LCFA\_INV\_COMPTON\_PHOTON\_energy} by giving $r_1$ as input parameter to sample the LCFA-predicted emitted photon energy $\bar{\gamma}_\gamma$. If $\bar{\gamma}_\gamma > \gamma_{\gamma,\text{LCFA}}^{(n)}$, the routine simply returns $\bar{\gamma}_\gamma$. Otherwise, \texttt{SFQED\_BLCFA\_INV\_COMPTON\_PHOTON\_energy} uses the second random number $r_2$ to evaluate the condition
\begin{equation} \label{rej_cond}
r_2 \left[\frac{d^2W_{pe}}{dtd\gamma_\gamma}(\bar{\gamma}_\gamma, \gamma_e, \chi_e) \right] \leq  \frac{d^2W_{pe}}{dtd\gamma_\gamma}(\gamma_{\gamma,\text{LCFA}}^{(n)}, \gamma_e, \chi_e).
\end{equation}
If Eq.~\eqref{rej_cond} is fulfilled, then the routine returns $\bar{\gamma}_\gamma$. Otherwise, it returns zero, which implies that no photon emission occurs. Note that in Eq.~\eqref{rej_cond} $\gamma_e$ has been written for clarity but is a global constant factor which multiplies both sides therefore canceling out. Hence, it is not used for evaluating Eq.~\eqref{rej_cond} in the code. In addition, for its implementation the change of variable $\gamma_\gamma = (3 \gamma_e \chi_e w^3)/(2 + 3 \chi_e w^3)$ was performed, which gives
\begin{equation} \label{chang}
\frac{d^2W_{pe}}{dt d\gamma_\gamma} = \frac{d^2W_{pe}}{dt dw} \left(\frac{d\gamma_\gamma}{dw}\right)^{-1} = \frac{d^2W_{pe}}{dt dw} \frac{(2 + 3 w^3 \chi_e)^2}{18 \gamma_e w^2 \chi_e}.
\end{equation}
The above change of variable is motivated by the fact that the function $d^2W_{pe} / dtdw$ is efficiently approximated with Chebyshev polynomials in SFQEDtoolkit. By contrast, $d^2W_{pe}/dt d\gamma_\gamma$ exhibits a $\gamma_\gamma^{-2/3}$ divergence when $\gamma_\gamma$ tends to zero, which makes it unsuitable for being expanded with Chebyshev polynomials, directly (see Sec.~\ref{impl}).

\begin{figure}[!h]
\centering
\includegraphics[width=1.\linewidth]{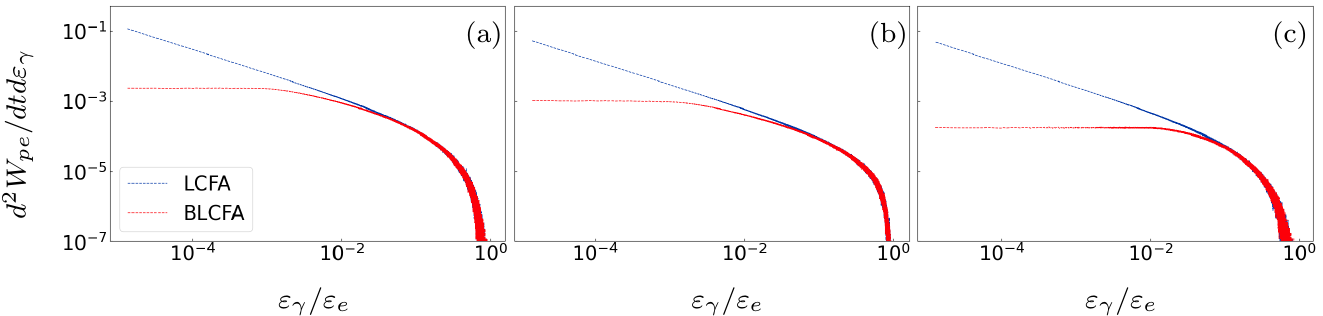}
\caption{The LCFA vs the beyond LCFA (see Ref.~\cite{dipiazzaPRA19}) differential photon emission probability for an electron colliding head-on with a plane-wave pulse for the following parameters: (a) 5~GeV electron initial energy, 5~fs (FWHM of the intensity) pulse duration, and $a_0 = 8$ normalized field amplitude; (b) 10~GeV electron initial energy, 5~fs pulse duration, and $a_0 = 10$ normalized field amplitude; (c) 10~GeV electron initial energy, 10~fs pulse duration, and $a_0 = 3$ normalized field amplitude.} 
\label{blcfa_fig}
\end{figure}

Figure~\ref{blcfa_fig} displays the results obtained after implementing SFQEDtoolkit routines into a fourth-order Runge-Kutta pusher, and by repeating the electron beam-laser pulse simulations with the same choice of electron and laser parameters as in Ref.~\cite{dipiazzaPRA19}. By comparing the results in Fig.~\ref{blcfa_fig} with the corresponding results in Ref.~\cite{dipiazzaPRA19}, it is apparent that the improved beyond LCFA spectrum is recovered.
\captionsetup{skip=3pt}

\section{Conclusions and outlook} \label{concl}

In this work we have presented a novel approach that allows an efficient implementation of the complex and computationally expensive functions needed to model strong-field QED processes into codes. This approach leverages on a combination of advanced function approximation techniques, and its key concepts can be naturally used in other areas of research. The method resorts to a combination of: (i) function approximation with Chebyshev polynomials; (ii) asymptotic expansions; (iii) variable and function transformation (see Sec.~\ref{impl}). We have applied this method to create an open source library named SFQEDtoolkit, which is designed to allow users for a straightforward implementation of SFQED processes, namely nonlinear Compton emission and nonlinear Breit-Wheeler pair creation, in existing particle-in-cell and Monte Carlo codes.

SFQEDtoolkit provides users with an efficient and better than 0.1\% accuracy implementation of SFQED processes throughout the whole particles' spectrum. Benchmarks performed with the PIC code Smilei version 4.7 have shown that SFQEDtoolkit outperforms the default 256-points tables and achieves an accuracy better than that of 1024-points lookup-tables. Currently, photon emission and pair creation with $\chi_{e/\gamma} \leq 2000$ are implemented by assuming the locally-constant-field approximation and collinear emission of the generated particles. For photon emission, the more advanced method beyond the locally-constant-field approximation presented in Ref.~\cite{dipiazzaPRA19} is also included. The implementation of the angular distribution of generated particles as well as of the spin and polarization-dependent SFQED processes is currently underway and \change{will be presented} in a separate publication.

\section{Acknowledgments}
This article comprises parts of the Ph.D. thesis work of Samuele Montefiori, to be submitted to the Heidelberg University, Germany.

\appendix

\section{User guide} \label{user}

SFQEDtoolkit is an open source library written in \texttt{C++}. A wrapper for \texttt{Fortran}-based programs which leverages the standardized interoperability of modern \texttt{Fortran} with \texttt{C} is also provided with the current version. SFQEDtoolkit can be used as a black box by following the instructions below and the examples showing its usage provided on GitHub at:\\
\url{https://github.com/QuantumPlasma/SFQEDtoolkit}.

This appendix describes the \change{key steps needed for implementing SFQEDtoolkit, and the main functions} currently available. SFQEDtoolkit exposes all functions\footnote{\change{Possible other minor functions, typically variation of the main functions, are included to address specific requests of some users.}} to the user through the module \texttt{SFQEDtoolkit\_Interface.hpp} and \texttt{SFQEDtoolkit\_Interface.f90} to \texttt{C++} and \texttt{Fortran} codes, respectively. We recommend users to read this appendix before implementing the library.

\change{SFQEDtoolkit must be initialized and finalized once at the beginning and at the end of the simulation, respectively. These are the only functions that must be implemented in a code using SFQEDtoolkit. All other available SFQEDtoolkit functions are independent of each other and can be ignored if not required, i.e., only the functions that are desired have to be implemented in a code. Thus, for example, none of the functions of the beyond LCFA method of Sec.~\ref{blcfa} needs to be implemented if not used. Analogously, for example, if one is interested in photon emission according to the LCFA method but not in pair creation, then the implementation of the function that compute the LCFA photon emission rate and of the function that calculate the LCFA emitted photon energy \change{}suffice. In fact, e.g., for testing purposes, one can even implement only the SFQEDtoolkit function for the LCFA photon emission rate, and use a custom version for calculating the LCFA emitted photon energy, or vice versa. This enables high flexibility and allows users to customize SFQEDtoolkit to their code and to their specific objective without implementing unnecessary functions.}

\begin{figure}[!t]
\centering
\includegraphics[width=1.\linewidth]{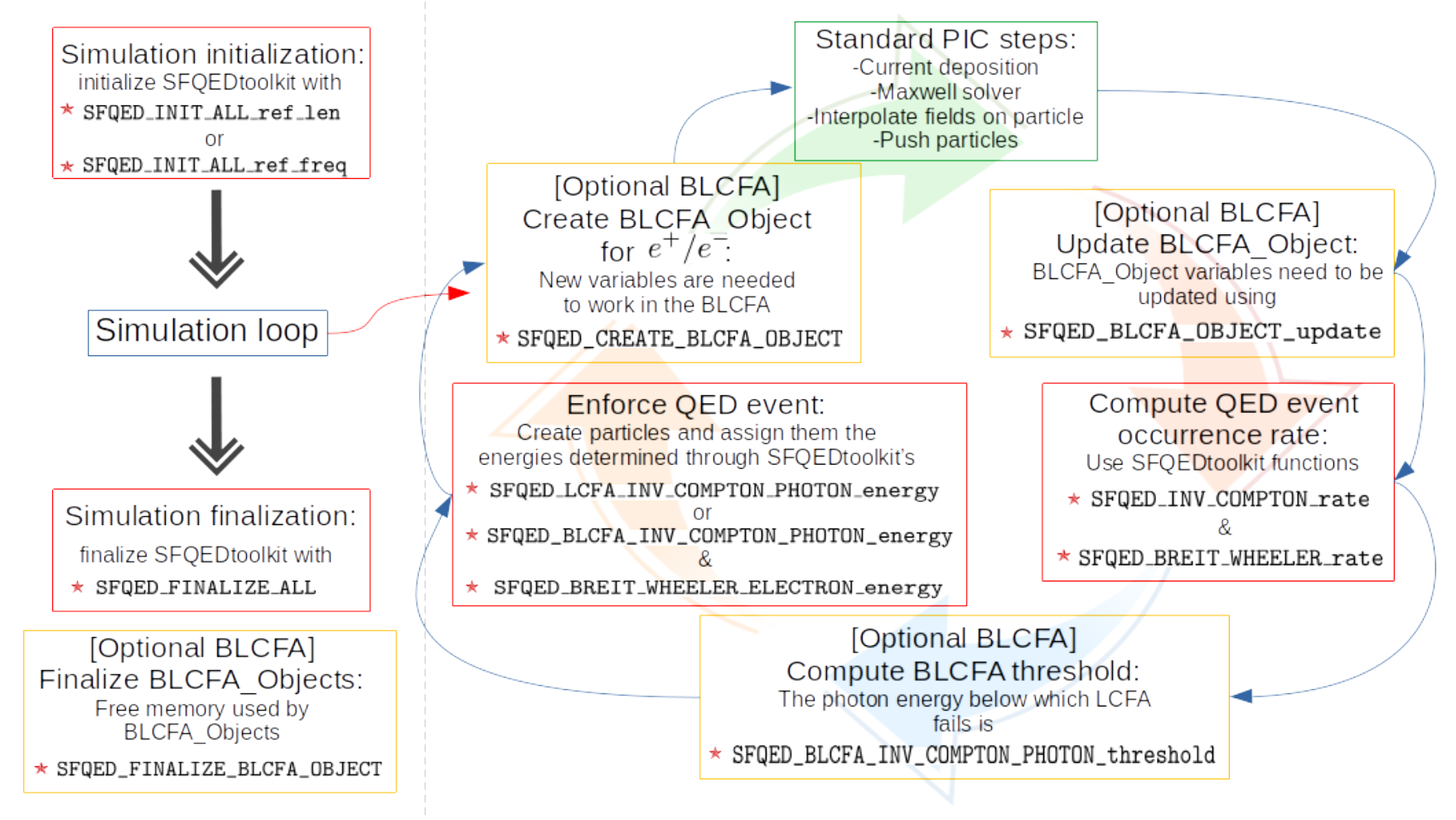}
\caption{Workflow of a PIC code implementing the routines provided by SFQEDtoolkit.}
\label{picloop}
\end{figure}

Figure \ref{picloop} displays the flowchart of a PIC code embedding SFQEDtoolkit. At initialization all the precomputed Chebyshev coefficients stored in the \texttt{txt} files included with the library are loaded into memory\footnote{\change{The tables with the Chebyshev coefficients are available in the ``coefficients'' folder of SFQEDtoolkit at \url{https://github.com/QuantumPlasma/SFQEDtoolkit}. The ``coefficients'' folder includes a ``README.md'' file detailing how to interpret the content of each file, and how the Chebyshev coefficients stored in the files are connected to the function that they approximate, including the specific interval of approximation.}}. At this stage the user needs to specify either the reference length or the reference frequency of the simulation. This is required because in SFQEDtoolkit all quantities are given in normalized units, and an angular frequency $\omega_r$ is used as a reference. Consequently, the reference time $T_r=1/\omega_r$, length $\lambda_r=c/\omega_r$, and field $E_r=m_e c \omega_r/|e|$ are obtained, while energy is normalized to $m_e c^2$. In particular, notice that in a problem where lengths are given in units of the laser wavelength $\lambda$, then $\lambda_r = \lambda / 2 \pi$ such that $\omega_r = 2 \pi c/ \lambda$. 
\captionsetup{skip=3pt}

The initialization is carried out through either the first or the second of the two following functions
\begin{nfunc}{func:one}
\texttt{bool} \texttt{SFQED\_INIT\_ALL\_ref\_len} \texttt{double} \texttt{ref\_len,} \texttt{double} \texttt{ts)}: initializes the environment by loading into memory all coefficients needed for modeling SFQED processes. In addition, it stores the reference length \texttt{ref\_len}$=\lambda_r$ and the timestep of the simulation \texttt{ts}$=\Delta t$. The reference length must be given in meters, while the simulation timestep must be given in normalized units, i.e., in units of $T_r = 1/\omega_r$. The simulation timestep is used only by the routines that implement the beyond LCFA photon emission method described in Sec.~\ref{blcfa}. \change{However, because of its generality, an input value is expected also when the beyond LCFA photon emission functions are not implemented}. At runtime, SFQEDtoolkit reads the required information from files located in the subdirectory ``coefficients'' from the parent directory where the program is executed. If initialization is successful, the function returns a boolean that is \texttt{true}.
\end{nfunc}
\begin{nfunc}{func:two}
\texttt{bool} \texttt{SFQED\_INIT\_ALL\_ref\_freq(} \texttt{double} \texttt{ref\_freq,} \texttt{double} \texttt{ts)}: equivalent and alternative to function \funcref{func:one}. The only difference is that the scale of the simulation is set by providing the reference angular frequency \texttt{ref\_freq}$=\omega_r$ in SI units.
\end{nfunc}

If the photon emission method beyond LCFA of Sec.~\ref{blcfa} is employed, SFQEDtoolkit provides users with a \texttt{C++} object named \texttt{BLCFA\_Object} that is designed to allocate the additional information that is required to keep track of the force acting on the particle. Namely, a boolean signaling whether the particle was created at the penultimate timestep, and two arrays each with three double precision elements. One array stores the transverse Lorentz force at the penultimate timestep, the other array stores the difference of the Lorentz force between the penultimate and the antipenultimate timestep. This object is created by calling the function 
\begin{nfunc}{func:three}
\texttt{BLCFA\_Object*} \texttt{SFQED\_CREATE\_BLCFA\_OBJECT()}: creates an object designed to store the information required to apply the beyond LCFA method to a particle. The boolean is set to \texttt{true}, while the two arrays are initialized to zero. The function returns a pointer to the created \texttt{BLCFA\_Object}.
\end{nfunc}
\change{Note that functions that allow users to implement the beyond LCFA method without employing a \texttt{BLCFA\_Object} are also provided and described below.} \change{If function \funcref{func:three} is implemented,} each computational particle should have one \texttt{BLCFA\_Object} associated. It is up to the user to organize objects in a suitable data structure (array, linked list, etc.) \change{and, if the code in which SFQEDtoolkit is implemented uses message passing interface (MPI), to manage the communication of the object data among MPI domains}. A recommended choice is to use inheritance to derive from \texttt{BLCFA\_Object} an object \texttt{Particle} where information such as particle position and momentum is stored, and to create either an array or a \texttt{C++} vector to store the data of all particles in the simulation. Another option is to include a pointer to a \texttt{BLCFA\_Object} in the class or data type used by the existing code to store information about a particle's status. \change{In a parallel code using MPI, the communication of the additional particle data stored in} \texttt{BLCFA\_Object} \change{should follow the same methodology used by the code to communicate other particle information such as particle's position and momentum. In fact, one should simply communicate the elementary datatypes stored in a} \texttt{BLCFA\_Object} \change{instead of the} \texttt{BLCFA\_Object} itself.

Next, the standard steps of a PIC loop are performed: currents are deposited on the grid, Maxwell equations are solved, the resulting fields on the grid are interpolated at the particle's position, and particles are advanced in time according to the Lorentz force with a suitable particle `pusher' (see, e.g., Ref.~\cite{birdsall-langdon}). If a leapfrog integrator is used, a convenient choice is to adapt the implementation to its staggered method by first advancing the particle momentum, then call the routines for modeling SFQED processes, and only afterward advance the particle position and store the new data into memory.

Again, only when using the beyond LCFA method for photon emission of Sec.~\ref{blcfa}, a user employing the \texttt{BLCFA\_Object} and its features can resort to the function
\begin{nfunc}{func:four}
\texttt{bool} \texttt{SFQED\_BLCFA\_OBJECT\_update(} \texttt{BLCFA\_Object*} \texttt{obj,} \texttt{double*} \texttt{p\_push,} \texttt{double*} \texttt{p,} \texttt{double} \texttt{delta,} \texttt{double} \texttt{gamma,} \texttt{double} \texttt{chi)}: if the boolean of the \texttt{BLCFA\_Object} pointed by \texttt{obj} is \texttt{true}, calculates $\bm{F}_{L,\perp}^{(n)}$ from Eqs.~\eqref{fl}-\eqref{fl_perp} by using the particle momentum before \texttt{p} and after \texttt{p\_push} one timestep, and changes the boolean of \texttt{BLCFA\_Object} to \texttt{false}. In addition, it sets both arrays of the \texttt{BLCFA\_Object} equal to the computed $\bm{F}_{L,\perp}^{(n)}$, while the boolean returned by the function is \texttt{false} to signal no emission at this timestep. By contrast, if the boolean of the \texttt{BLCFA\_Object} pointed by \texttt{obj} is \texttt{false}, uses the particle momentum before \texttt{p} and after \texttt{p\_push} one timestep to perform all calculations in Eqs.~\eqref{fl}-\eqref{condit}. Then, it returns \texttt{delta} as defined in Eq.~\eqref{condit}, the normalized particle energy \texttt{gamma} [Eq.~\eqref{gammaBLCFA}], and its quantum parameter \texttt{chi} [Eq.~\eqref{chi_p}]. Finally, the two arrays of the \texttt{BLCFA\_Object} pointed by \texttt{obj} are updated by storing $\bm{F}_{L,\perp}^{(n)}$ and $\bm{F}_{L,\perp}^{(n)}-\bm{F}_{L,\perp}^{(n-1)}$, while the boolean returned by the function is \texttt{true} only if the necessary conditions for applying the local emission model hold [see the conditions below Eq.~\eqref{condit}].
\end{nfunc}

While functions \funcref{func:three} and \funcref{func:four} are relevant only for the photon emission model beyond LCFA of Sec.~\ref{blcfa}, SFQED probability rates must be computed for each particle and at each timestep in the considered LCFA and beyond LCFA models. SFQEDtoolkit allows users to efficiently compute the SFQED rates by means of
\begin{nfunc}{func:five}
\texttt{double} \texttt{SFQED\_INV\_COMPTON\_rate(} \texttt{double} \texttt{gamma,} \texttt{double} \texttt{chi)}: uses the normalized electron or positron energy \texttt{gamma} as well as its quantum parameter \texttt{chi} to return the LCFA photon emission probability rate.
\end{nfunc}
\begin{nfunc}{func:six}
\texttt{double} \texttt{SFQED\_BREIT\_WHEELER\_rate(} \texttt{double} \texttt{gamma,} \texttt{double} \texttt{chi)}: uses the normalized photon energy \texttt{gamma} as well as its quantum parameter \texttt{chi} to return the LCFA probability rate of photon conversion in an electron-positron pair.
\end{nfunc}
\begin{nfunc}{func:seven}
\texttt{double} \texttt{SFQED\_BREIT\_WHEELER\_rate\_fast(} \texttt{double} \texttt{gamma,} \texttt{double chi)}: same as function \funcref{func:six} but it returns zero if \texttt{chi}$<0.3$ (see Sec.~\ref{pa_pr_rate}).
\end{nfunc}
We recall that, as thoroughly discussed in Sec.~\ref{probabs}, in the LCFA model both the local probability and the optical depth method can be used to determine whether a SFQED event is deemed to occur from probability rates. However, the beyond LCFA photon emission method of Sec.~\ref{blcfa} is based on an acceptance-rejection technique applicable only with the local probability method. 

In case a SFQED event is deemed to occur according to the LCFA probability rate, the energy of generated particles is computed by calling
\begin{nfunc}{func:nine}
\texttt{double} \texttt{SFQED\_LCFA\_INV\_COMPTON\_PHOTON\_energy(} \texttt{double} \texttt{gamma,} \texttt{double} \texttt{chi,} \texttt{double} \texttt{rnd)}: returns the normalized photon energy by sampling from the  LCFA energy distribution. Its input parameters are the normalized electron or positron energy \texttt{gamma}, the quantum parameter \texttt{chi} of the emitting particle, and a random number \texttt{rnd} sampled from a uniform distribution in $(0,1)$.
\end{nfunc}
\begin{nfunc}{func:eleven}
\texttt{double} \texttt{SFQED\_BREIT\_WHEELER\_ELECTRON\_energy(} \texttt{double} \texttt{gamma,} \texttt{double} \texttt{chi,} \texttt{double} \texttt{rnd)}: returns the normalized energy of the generated electron (positron) by sampling from the LCFA energy distribution for photon conversion into an electron-positron pair. The energy of the positron (electron) is then readily obtained from the difference between the photon and the electron (positron) energy. The normalized photon energy \texttt{gamma} and its quantum parameter \texttt{chi}  as well as a uniformly distributed random number \texttt{rnd} in $(0,1)$ must be passed as input parameters.
\end{nfunc}
\begin{nfunc}{func:twelve}
\texttt{double} \texttt{SFQED\_BREIT\_WHEELER\_ELECTRON\_energy\_fast(} \texttt{double} \texttt{gamma,} \texttt{double} \texttt{chi,} \texttt{double} \texttt{rnd)}: same as function~\funcref{func:eleven}, but \texttt{chi} $<0.3$ is not managed. This function must be used only in combination with function~\funcref{func:seven}.
\end{nfunc}

When the beyond LCFA model of Sec.~\ref{blcfa} is used, if function~\funcref{func:four} returns \texttt{true} and a SFQED event is deemed to occur according to the LCFA probability rate (see function~\funcref{func:five}), the calculation of the local LCFA energy threshold $\gamma_{\gamma,\text{LCFA}}$ as defined in Eq.~\eqref{gammaLCFA} is required to determine the improved photon energy distribution. If \texttt{BLCFA\_Object}s are created and updated through functions \funcref{func:three} and \funcref{func:four}, respectively, then $\gamma_{\gamma,\text{LCFA}}$ can be computed by calling
\begin{nfunc}{func:eight}
\texttt{double} \texttt{SFQED\_BLCFA\_INV\_COMPTON\_PHOTON\_threshold(} \texttt{BLCFA\_Object*} \texttt{obj,} \texttt{double} \texttt{delta,} \texttt{double} \texttt{gamma,} \texttt{double} \texttt{chi)}: returns the normalized photon energy threshold below which the LCFA breaks, i.e., the $\gamma_{\gamma,\text{LCFA}}$ defined in Eq.~\eqref{gammaLCFA}. The arguments of this routine are those passed to and updated by function~\funcref{func:four}.
\end{nfunc}
After calling function~\funcref{func:eight} all required quantities for the beyond LCFA method are available, and the photon energy is obtained from
\begin{nfunc}{func:ten}
\texttt{double} \texttt{SFQED\_BLCFA\_INV\_COMPTON\_PHOTON\_energy(} \texttt{double} \texttt{limit,} \texttt{double} \texttt{gamma,} \texttt{double} \texttt{chi,} \texttt{double} \texttt{rnd,} \texttt{double} \texttt{rnd2)}: returns the normalized photon energy according to the beyond LCFA model of Sec.~\ref{blcfa}. This function requires as input the same parameters of function~\funcref{func:nine}, plus two additional arguments: (i) the \texttt{limit=}$\gamma_{\gamma,\text{LCFA}}$ defined in Eq.~\eqref{gammaLCFA}, which users can obtain by calling function~\funcref{func:eight}, and (ii) \texttt{rnd2}, which is a uniformly distributed random number in $(0,1)$ independent of \texttt{rnd}. If $\gamma_{\gamma,\text{LCFA}}>0.75\gamma_e$ the function returns zero. Otherwise, it applies the acceptance-rejection technique of Sec.~\ref{blcfa} and returns zero if the event is rejected. 
\end{nfunc}

After the above steps, the computational cycle is complete and is repeated until the end of the simulation is reached. At this point, if the \texttt{BLCFA\_Object} was used, memory should be deallocated by calling the function
\begin{nfunc}{func:thirteen}
\texttt{void} \texttt{SFQED\_FINALIZE\_BLCFA\_OBJECT(} \texttt{BLCFA\_Object* obj)}: deallocates the memory space reserved for the \texttt{BLCFA\_Object} pointed by \texttt{obj}.
\end{nfunc}
and SFQEDtoolkit should be `finalized', i.e.,  the memory reserved by SFQEDtoolkit for storing the tables of coefficients should be deallocated, by calling the function
\begin{nfunc}{func:fourteen}
\texttt{void} \texttt{SFQED\_FINALIZE\_ALL()}: frees all the memory allocated at initialization.
\end{nfunc}

\change{The BLCFA routines \funcref{func:four} and \funcref{func:eight} both rely on the \texttt{BLCFA\_Object} entity, which is initialized with function \funcref{func:three}. If desired, the user can completely bypass the usage of \texttt{BLCFA\_Object}s. In this case, the declaration and initialization of the additional variables that are necessary for the BLCFA algorithm has to be done by the user, directly, and functions \funcref{func:four} and \funcref{func:eight} are replaced by
\begin{nfunc}{func:rawupdate}
\texttt{bool} \texttt{SFQED\_BLCFA\_OBJECT\_update\_raw(} \texttt{double*} \texttt{p\_push,} \texttt{double*} \texttt{p,} \texttt{double*} \texttt{Lorentz\_F,} \texttt{double*} \texttt{Delta\_Lorentz\_F,} \texttt{bool} \texttt{just\_created,} \texttt{double} \texttt{delta,} \texttt{double} \texttt{gamma,} \texttt{double} \texttt{chi)}: replaces function \funcref{func:four}. Namely, if \texttt{just\_created} is \texttt{true}, calculates $\bm{F}_{L,\perp}^{(n)}$ from Eqs.~\eqref{fl}-\eqref{fl_perp} by using the particle momentum before \texttt{p} and after \texttt{p\_push} one timestep, and changes \texttt{just\_created} to \texttt{false}. In addition, it sets \texttt{Lorentz\_F} and \texttt{Delta\_Lorentz\_F} equal to the computed $\bm{F}_{L,\perp}^{(n)}$, and the boolean returned by the function is \texttt{false} to signal no emission at this timestep. 
Note that for a new particle, the user must have declared in the code suitable variables corresponding to \texttt{Lorentz\_F} and \texttt{Delta\_Lorentz\_F}, both set to zero, and to \texttt{just\_created}, set to \texttt{true}. By contrast, if \texttt{just\_created} is \texttt{false}, uses the particle momentum before \texttt{p} and after \texttt{p\_push} one timestep to perform all calculations in Eqs.~\eqref{fl}-\eqref{condit}. Then, it returns \texttt{delta} as defined in Eq.~\eqref{condit}, the normalized particle energy \texttt{gamma} [Eq.~\eqref{gammaBLCFA}], and its quantum parameter \texttt{chi} [Eq.~\eqref{chi_p}]. Finally, \texttt{Lorentz\_F} and \texttt{Delta\_Lorentz\_F} are updated and store $\bm{F}_{L,\perp}^{(n)}$ and $\bm{F}_{L,\perp}^{(n)}-\bm{F}_{L,\perp}^{(n-1)}$, respectively, while the boolean returned by the function is \texttt{true} only if the necessary conditions for applying the local emission model hold [see the conditions below Eq.~\eqref{condit}].
\end{nfunc}
and
\begin{nfunc}{func:rawthresh}
\texttt{double} \texttt{SFQED\_BLCFA\_INV\_COMPTON\_PHOTON\_threshold\_raw(} \texttt{double*} \texttt{Lorentz\_F,} \texttt{double} \texttt{delta,} \texttt{double} \texttt{gamma,} \texttt{double} \texttt{chi)}: replaces function \funcref{func:eight}. Namely, it returns the normalized photon energy threshold below which the LCFA breaks, i.e., the $\gamma_{\gamma,\text{LCFA}}$ defined in Eq.~\eqref{gammaLCFA}. The arguments of this routine are those passed to and updated by function~\funcref{func:rawupdate}.\end{nfunc}
}
Finally, SFQEDtoolkit provides users with two functions for the computation of the particle quantum nonlinearity parameter $\chi$ according to Eq.~\eqref{chi_exact}, and one function implementing the collinear emission model for the generated particle. 
\begin{nfunc}{func:fifteen}
\texttt{double} \texttt{compute\_chi\_with\_vectors(} \texttt{double} \texttt{gamma,} \texttt{double} \texttt{p[3],} \texttt{double} \texttt{EE[3],} \texttt{double} \texttt{BB[3])}: returns the quantum nonlinearity parameter $\chi_{\gamma/e}$ of a particle with energy \texttt{gamma} and momentum \texttt{p} in an electric \texttt{EE} and magnetic \texttt{BB} field. All quantities must be provided in normalized units.
\end{nfunc}
\begin{nfunc}{func:sixteen}
\texttt{double} \texttt{compute\_chi\_with\_components(} \texttt{double} \texttt{gamma,} \texttt{double} \texttt{px,} \texttt{double} \texttt{py,} \texttt{double} \texttt{pz,} \texttt{double} \texttt{EEx,} \texttt{double} \texttt{EEy,} \texttt{double} \texttt{EEz,} \texttt{double} \texttt{BBx,} \texttt{double} \texttt{BBy,} \texttt{double} \texttt{BBz)}: same as \funcref{func:fifteen} but the electric and magnetic field are passed component-by-component.
\end{nfunc}
\begin{nfunc}{func:seventeen}
\texttt{void} \texttt{SFQED\_build\_collinear\_momentum(} \texttt{double} \texttt{gamma\_out,} \texttt{double} \texttt{p\_in[3],} \texttt{double} \texttt{p\_out[3])}: returns the momentum of the generated particle \texttt{p\_out} aligned with the momentum of the parent particle \texttt{p\_in} and sets its magnitude according to the energy of the generated particle \texttt{gamma\_out}. All parameters are in normalized units.
\end{nfunc}
The above functions can be used, e.g., in combination with functions~\funcref{func:five}--\funcref{func:twelve}.

\section{Clenshaw's recurrence formula} \label{clensh}

Once the coefficients $c_i$ of the Chebyshev expansion are known, one may directly compute the function
\begin{equation}
f(x) = \sum_{i=0}^N c_i T_i(x) = c_0T_0(x) + c_1T_1(x) + ... + c_NT_N(x),
\label{f_brutal}
\end{equation}
by evaluating the Chebyshev polynomials, e.g., via their defining recurrence relation in Eq.~\eqref{cheb_rec_rel}, and then sum their contributions. However, there are delicate cancellations of terms in the sum, and first computing the polynomials and then summing contributions is not stable against round-off errors. By contrast, Clenshaw's recurrence provides a numerically stable and efficient approach to evaluate Eq.~\eqref{f_brutal}.
By starting from the Chebyshev coefficients $c_i$, Clenshaw's recurrence exploits the recurrence relation of Eq.~\eqref{cheb_rec_rel}, which can be recast as
\begin{equation} \label{cheb_rel_rew}
T_{n}(x) - 2xT_{n+1}(x) + T_{n+2}(x) = 0,
\end{equation}
to build a new set of `coefficients' $b_i$ defined as
\begin{equation} \label{back_rec}
b_{n} - 2 x b_{n+1} + b_{n+2} = c_n,\qquad b_{N+1}=b_{N+2}=0.
\end{equation}
Notice that, although not explicitly indicated, the definition of the $b_i$ via Eq.~\eqref{back_rec} implies that the $b_i$ depend on $x$, in general. By plugging Eq.~\eqref{back_rec} into Eq.~\eqref{f_brutal}, we get
\change{
\begin{align}
f(x) &= (b_{0} - 2xb_{1} + {\color{black}b_{2}})T_0(x) + \notag\\
&\,+ (b_{1}  {\color{black}- 2xb_{2}} +  {\color{black}b_{3}})T_1(x) +\notag\\
&\,+ ( {\color{black}b_{2}} {\color{black}- 2xb_{3}} +   {\color{black}b_{4}})T_2(x) +\notag\\
&\,+ ... +\notag\\
&\,+ ( {\color{black}b_{N-3}}  {\color{black}- 2xb_{N-2}} +  {\color{black}b_{N-1}})T_{N-3}(x) +\notag\\
&\,+ ( {\color{black}b_{N-2}}  {\color{black}- 2xb_{N-1}} +  {\color{black}b_{N}})T_{N-2}(x) +\notag\\
&\,+ ( {\color{black}b_{N-1}}  {\color{black}- 2xb_{N}})T_{N-1}(x) +\notag\\
&\,+  {\color{black}b_N}T_N(x),
\label{f_brutal_1}
\end{align}
}
where the \change{}diagonal terms cancel out by virtue of Eq.~\eqref{cheb_rel_rew}. Now, Eq.~\eqref{f_brutal_1} simplifies to
\begin{align}
f(x) &= \sum_{i=0}^{N}c_iT_i(x)=\notag\\
&= b_{0}T_0(x) - 2xb_{1}T_0(x) + b_{1}T_1(x) =\notag\\
&= b_{0} - 2xb_{1} + b_{1}x =\notag\\
&= b_{0} - xb_{1}.
\label{f_brutal_2}
\end{align}
Thus, there is no need to compute the Chebyshev polynomials $T_i(x)$. Elegant and computationally efficient, Clenshaw's recurrence is also stable against round-off error, as reported in Ref.~\cite{pressNR3rd}:
\begin{displayquote}
You need to be aware that recurrence relations are not necessarily stable against roundoff error in the direction that you propose to go (either increasing $n$ or decreasing $n$). [...] Clenshaw's recurrence is always stable, independent of whether the recurrence for the functions $F_k$ [the functions appearing in the series] is stable in the upward or downward direction.
\end{displayquote}
The same procedure can be used for multivariate functions. For instance, if
\begin{equation} \label{start_cl}
f(x,y) = \sum_{i=0}^{N}\Bigl[\sum_{j=0}^{M}c_{ij}T_{j}(y)\Bigr] T_{i}(x)
\end{equation}
is the truncated Chebyshev expansion of a two-variable function, we can first apply Clenshaw's recurrence to build a set of $b_{i,(m)}$ from $c_{ij}$ by keeping the index $i$ fixed [see the terms that have been grouped within the square brackets of Eq.~\eqref{start_cl}]
\begin{equation} \label{back_rec_2d}
\forall i, \qquad
b_{i,(m)} - 2yb_{i,(m+1)} + b_{i,(m+2)} = c_{im}, \qquad
b_{i,(M+1)}=b_{i,(M+2)}=0,
\end{equation}
and get $N+1$ coefficients $g_i(y) = b_{i,(0)} - y b_{i,(1)}$. Now, Eq.~\eqref{start_cl} becomes
\begin{equation} \label{start_cl_2}
f(x,y) = \sum_{i=0}^{N}\Bigl[b_{i,(0)} - yb_{i,(1)}\Bigr] T_{i}(x) \equiv \sum_{i=0}^{N}g_i(y) T_{i}(x).
\end{equation}
Finally, the application of Clenshaw's recurrence to the coefficients $g_i(y)$ gives $f(x,y)$ evaluated at the desired $x,y$. Remarkably, if $f(x,y)$ needs to be evaluated multiple times and $y$ i\change{s} fixed, \change{then} the $N+1$ coefficients $g_i(y)$ do not change therefore greatly reducing the computational cost of the further evaluations. The above iterative procedure can be naturally extended to functions of an arbitrary number of variables.

\bibliographystyle{elsarticle-num}
\hyphenation{Post-Script Sprin-ger}

\end{document}